\documentstyle[psfig]{aa}

\input epsf         
\begin{document}
\title{Escape probability methods versus ``exact" transfer for modelling 
the X-ray spectrum 
of Active Galactic Nuclei and X-ray binaries}

\author{Anne-Marie Dumont\inst{1}, Suzy Collin\inst{1}, Fr\'ed\'eric 
Paletou\inst{2},\\ S\'everine 
Coup\'e\inst{1,3}, Olivier Godet\inst{4}, Didier Pelat\inst{1}}

\offprints{Anne-Marie Dumont (anne-marie.dumont@obspm.fr)}

\institute{$^1$LUTH, Observatoire de Paris, Section de
Meudon, F-92195 Meudon Cedex, France\\
$^2$Observatoire de la C\^ote d'Azur, Dept. Cassini, BP 4229, 06304 
Nice Cedex 4, France\\
$^3$ Universit\'e Claude Bernard, 69000 Lyon, France\\
$^4$CESR,  9 av. du Colonnel Roche,31028 Toulouse
Cedex 4, France}
\date{Received 6 February 2003 : / Accepted 27 May 2003 : }

\titlerunning{Modelling the spectrum of AGN and X-ray binaries}
\authorrunning{A-M. Dumont et al.}

\abstract{In the era of XMM-Newton and Chandra missions, it is crucial 
to use 
codes able to compute correctly the line spectrum  
of X-ray irradiated thick media (Thomson thickness of the order of 
unity), in order to build models for 
the structure and the emission 
of the central regions of Active Galactic Nuclei (AGN), or of X-ray 
binaries. In all 
photoionized codes except in our code Titan, 
the line intensities are computed with the 
so-called ``escape probability approximation".
In its last version,  Titan 
solves the 
transfer of a thousand lines and of the continuum with the ``Accelerated Lambda 
Iteration" method, which is one of the most efficient and at the same 
time the most secure for line transfer. We first review  
the escape probability formalism and mention various reasons why it 
should lead to wrong results concerning the 
line fluxes. Then we check several
approximations 
commonly 
used instead of line transfer in photoionization codes, by comparing them to the full
 transfer computation. We find that for conditions typical of
 the AGN or X-ray binary emission medium, all 
 approximations lead to an overestimation of the emitted 
X-ray line spectrum, which can reach more than one order of magnitude.
We show that it is due mainly to the local treatment of line 
photons, implying a delicate balance between excitations of X-ray
  transitions by the very intense underlying
 diffuse X-ray continuum (which are not taken properly into account in 
 escape probability approximations) and the net rate of excitations 
 by the diffuse line flux. The most affected lines are those in the 
 soft X-ray range.
  Such processes are much less important in 
 cooler and thinner media (like the Broad Line Region of AGN), as the most 
 intense lines lie in the optical and near ultraviolet range where
 the diffuse continuum is small.  We conclude
that it is very important to treat correctly the transfer
of the continuum to get the best results for the line 
 spectrum. On the other 
 hand the approximations used for the escape probabilities
have a relatively small 
 influence on the computed thermal and ionization structure of the
surface layers, but in the deep layers, they lead to an 
 overestimation of the ionization state. As a consequence the 
computed continuum emitted by the back (non-irradiated) side is not correct,
 and might 
be strongly overestimated in the EUV range.

}

\maketitle

\section{Introduction}

For about two decades, many computations of the structure and emission of 
irradiated hot media have been performed, to understand the 
spectra of Active 
Galactic Nuclei (AGN) or of X-ray binaries. Based on the observations,  
all interpretations of AGN 
postulate that their 
 UV-soft X spectrum is emitted by a dense, warm, and optically thick medium (density higher 
 than  10$^{12}$ cm$^{-3}$, temperature of 10$^5$ to 10$^6$ 
 K, Thomson thickness of the order of, or greater than, unity), illuminated by an X-ray 
 continuum. 
Similar conditions are met in X-ray binary stars.  
This medium can be identified with the atmosphere of an accretion 
disc in hydrostatic equilibrium (Ko \& Kallman 1994, 
Nayakshin et al. 2000, Ballantyne et al. 2001, 
 R\' o\. za\' nska et al. 
2002), or with a system of clouds, either of constant density, or of 
constant pressure (Collin-Souffrin et al. 1996, Dumont et al. 2002). In 
early models, the disc itself was assumed to be a slab of constant 
density (Ross \& Fabian 1993, and subsequent works). Whatever it is, this 
medium is assumed to be heated and photoionized by an X-ray 
continuum extending up to a few hundreds keV.

Similar conditions hold in the atmospheres of the discs in X-ray 
binaries: they are also dense and thick, and irradiated by a strong X-ray 
continuum. Their modelling is performed exactly like those of AGN 
discs (cf. Ko \& Kallman 1994).

Several codes have been developed in order to 
handle such computations. Three codes are particularly designed for Thomson thick 
media: Nayakshin's code (Nayakshin et 
al. 2000), Ross and Ballantyne's code (Ross et al. 1978, Ross 1979), and our code, Titan, 
described in Dumont et al. (2000), hereafter called DAC, updated in Dumont \& 
Collin (2001), in Coup\'e (2002), and finally in the present 
paper where a new transfer method for the lines (the Accelerated Lambda 
Iteration method) 
has been implemented. Nayakshin's code was especially 
developed for the irradiated
atmosphere of an accretion disc in hydrostatic equilibrium, and Ross code and
 Titan have 
been adapted to this case (cf. Ballantyne et al. 
2001, and  R\' o\. za\' nska et al. 2002). A comparison between 
these codes was performed in the case of a photoionized 
slab with a constant 
density, and the resulting spectra are shown in Pequignot et al. (2001). 
It is clear that they are all different, as a result of the 
approximations made in the computations, and in spite of the fact 
that the parameters of the model were exactly the same. One should 
also mention Hubeny's code (Hubeny et al. 2000, 2001), which 
has produced accurate grids of model spectra for accretion disc 
atmospheres, but 
is still not adapted to the X-ray irradiated case.

With the advent of the X-ray missions Chandra and XMM-Newton, high 
resolution spectra of AGN in the soft X-ray range have been obtained 
showing features in absorption and/or in emission. There is in particular a strong 
controversy on the identification of the lines in the spectrum of MCG -6-30-15. 
Branduardi-Raymont et al. (2001) propose that the spectrum is dominated 
by very intense lines in emission (for instance OVIII L$\alpha$ with an equivalent width of about
 100 eV)
 emitted by the atmosphere of an irradiated relativistic disc,
 while for Lee et al. (2001) 
the spectral features are mainly absorption edges and absorption lines 
of neutral elements 
locked in a dusty absorber.  It is therefore crucial to get good 
model spectra in order to check the validity of the first hypothesis, 
and more generally to be able to model the X-ray line spectrum. 

Several well-known and publicly available codes 
are also frequently used for photoionized media (Cloudy and XSTAR in particular).
 They are very accurate 
from the point of view of the atomic data and for the number of ions and 
discrete levels taken into account. In DAC it was shown that the transfer 
treatment of the 
continuum performed in these codes 
(outward only approximation, or escape probability formalism)
precludes them to be used for Thomson thick media. 
We will show that the same restriction holds for the line transfer.   

Titan, Ross-Ballantyne's and Nayakshin's codes 
are non-LTE photoionization codes differing both by the treatment of
 the atomic data and of the transfer. These points have been 
discussed in some details in  R\' o\. za\' nska et al. (2002) and in 
Dumont et al. (2002), so we will not insist on them here. It is sufficient 
to recall that, as long as atomic data are considered, Nayakshin's 
code, which is based on the code XSTAR, is the best. 
Titan is 
intermediate between Nayakshin and Ross-Ballantyne codes, as it includes all ions 
of the ten most abundant elements, and treats them as interlocked multi-level atoms 
in the case of all hydrogen-like, helium-like, and lithium-like species 
 and a few other ions. 

For the transfer 
of the continuum, 
 Ross-Ballantyne code uses the Kompaneets diffusion approximation, and Nayakshin 
 the variable Eddington factor transfer method with the addition of a
 Compton scattering operator, 
 while Titan uses  
the Accelerated Lambda Iteration (ALI, cf. Hubeny 2001, Dumont 
\& Collin 2001). Titan
considers the effect of Compton diffusions only in the energy balance, 
and has therefore to be coupled with the Monte Carlo 
code NOAR of DAC
to take into account Compton scatterings in the high energy continuum. 

For the transfer of the lines, Nayakshin's and 
Ross-Ballantyne's codes use the so-called
``escape probability approximation", like Cloudy and XSTAR. It means that 
 the line intensities, the contribution of the lines to the energy balance, 
 to the ionization equilibrium, and to the radiation pressure,
are 
computed in an approximate way, which might or might not give correct results.  
In this new version of Titan the 
 line transfer is solved now with
the ALI method, unlike to the previous version described in 
Dumont \& Collin (2001) where it was solved with a simple Lambda 
Iteration method (a full description of the present version of the code will 
be given elsewhere).
 This method ensures that even the most optically thick lines 
are accurately computed. Comptonisation of line photons was 
also recently added to the code as described 
in Appendix A\footnote{Actually it is taken into account in the line 
intensities but not in their profiles}.

A strong drawback of Titan is that it is time-consuming, and 
this prevents it from being easily used like the other codes, for instance
 to build grids of 
models. This is not only due to the resolution of the 
transfer in the lines, which requires to solve the transfer
for fifteen frequencies inside each line, for several hundreds
 of lines (presently more than 800, whose majority is optically thick)
  to be sure to have a good representation 
of the spectrum 
 and of the structure of the medium, but it is also due to the accurate 
 treatment of the continuum itself, which is necessary as we will see later. 
Thus 
 one could ask legitimely if it is worth performing these enormous 
computations, at the expense of other ones (for instance 
 calculating the equilibrium of a very large number of levels like in Cloudy 
or XSTAR). Note however that the computation time with the ALI method 
is reduced with respect to the previous transfer method, especially
using a diagonal i.e., local approximate operator as we did.

We will not discuss here the impact of the 
treatment of the continuum, as it was already done in DAC,  and we will
 focus only on the influence of the 
escape probability approximation when it is used instead of
line transfer. We 
will  thus compare  different line
 escape probability approximations with the full transfer
 treatment, keeping {\it exactly the same method for the transfer of the 
 continuum}. It means that we will not be able to compare directly our 
computations with the
Ross-Ballantyne and Nayakshin ones, since they do not use the same method for 
the transfer of the continuum. But doing that, we can disentangle the
influence of the line approximations from that of the treatment of the 
continuum, which is more important than the lines in defining the 
temperature and ionization
structure.

In this paper we restrict our study to two representative models. 
We will perform in a 
forthcoming paper the same
comparison for different physical conditions, in order to set limits 
to the validity of the escape approximations. 

 In the next section, we summarize the basics of radiation transfer and 
of the escape probability 
formalism. Some approximations used in 
different codes for the escape probabilities are recalled in Section 
3. In Section 4 
the structure and the emission spectrum of an illuminated slab of 
constant density  are 
computed for several approximations and compared to the results of the
 full 
transfer treatment. Section 5 is devoted to a discussion of the results.

\section{The transfer equation and the escape probability 
formalism}

 The transfer equation 
writes in a plane-parallel geometry: 

\begin{equation}
\mu {dI_{\nu} \over dz}= -(\kappa_{\nu} 
+\sigma) I_{\nu}+\sigma J_{\nu}+\epsilon_{\nu} \\
\label{eq-trans1}
\end{equation}
where $z$ is the distance to the illuminated edge,  $\mu$ is the cosine of 
the angle between the normal and the light ray, $I_{\nu}$ and $J_{\nu}$ 
are respectively the specific intensity in the direction $\mu$ and the 
angle averaged intensity at the frequency $\nu$, $\kappa_{\nu}$ is 
the absorption coefficient, $\sigma$ is the diffusion 
coefficient - here it is due to Thomson scattering and does not depend on 
the frequency - and $\epsilon_{\nu}$ is the emissivity.

This equation depends on $\mu$, on $\nu$, on $z$, and is therefore impossible to 
solve in its whole generality. The 
dependence in $\mu$ can be simplified by the use of a limited number of 
directions, for instance with a ``two-stream" approximation, as  in 
our code Titan.

Integrating Eq. \ref{eq-trans1} on the angles gives:

\begin{equation}
{dF_{\nu} \over dz}= - 4\pi\kappa_{\nu} J_{\nu}+4\pi \epsilon_{\nu} 
\label{eq-trans-angle}
\end{equation}
where $F_{\nu}$ is the flux at the frequency $\nu$. 

If we consider a frequency where a line is superposed on the 
continuum, one has
$\kappa_{\nu}= \kappa_{line}\phi_{\nu} + \kappa_{c}$, and 
$\epsilon_{\nu}= \epsilon_{line}\psi_{\nu} +\epsilon_{c}$,   
where $\kappa_{line}$ and $\epsilon_{line}$ are respectively the 
absorption coefficient and the emissivity integrated on the line 
profile, 
$\phi_{\nu}$ and $\psi_{\nu}$ are the absorption and emission 
normalized line profiles, $\kappa_{c}$ and $\epsilon_{c}$ are the 
absorption coefficient and the emissivity of the continuum at the 
line frequency. $\kappa_{line}$ and $\epsilon_{line}$ are given 
respectively by:

\begin{equation}
\kappa_{line}= {h\nu\over 4\pi} 
[ n_l B_{lu}- n_u B_{ul} ] 
\label{eq-kappa-line}
\end{equation}
and
\begin{equation}
\epsilon_{line}= {h\nu\over 4\pi} n_u A_{ul}
\label{eq-epsilon-line}
\end{equation}
where $A_{ul}$, $B_{ul}$ and $B_{lu}$ are the 
usual radiative (Einstein) excitation and deexcitation 
coefficients between the upper ($u$) and lower ($l$) levels of the 
line transition, and $n_u$ and $n_l$ are the numerical densities 
of the upper and lower levels.

Separating the flux of the line and of the underlying continuum, one gets:

\begin{equation}
{d(F_{\nu}-F_{c}) \over dz}=  4\pi 
\left[-\kappa_{line}\phi_{\nu}J_{\nu}
- \kappa_{c} (J_{\nu}-J_{c}) + 
\epsilon_{line}\psi_{\nu}\right] 
\label{eq-flux-line}
\end{equation}
where $J_{c}$ and $F_{c}$ are respectively the angle averaged intensity  
and the flux in the continuum, both at the line frequency. 

Integrating now on frequencies, one gets:

\begin{eqnarray}
{dF_{line} \over dz}={d(F-F_{c}) \over dz}&=&  4\pi 
[-\kappa_{line}\int \phi_{\nu}J_{\nu}d\nu \\ \nonumber  
&-&  \kappa_{c} J_{line} + \epsilon_{line}]
\label{eq-flux-line-integre}
\end{eqnarray}
where $F_{line}$ is the flux integrated on the line 
profile and $J_{line}=\int  (J_{\nu}-J_{c})d\nu$. This 
equation can also be written, using Eqs. \ref{eq-kappa-line} and
 \ref{eq-epsilon-line}:
 
 \begin{eqnarray}
{dF_{line} \over dz}&=&  h\nu 
[ n_u  B_{ul}\int J_{\nu}{\psi}_{\nu}d\nu - n_l 
B_{lu}\int J_{\nu}{\phi}_{\nu} d\nu ] \\ \nonumber  
&& + h\nu  n_u A_{ul}- 4\pi \kappa_{c} J_{line}. 
\label{eq-flux-line-integrebis}
\end{eqnarray}

It is important for the following to notice that $J_{\nu}$ is the 
total mean intensity including the continuum $J_{c}$.

It is useful to introduce the source function $S_{\nu}$:

\begin{equation}
S_{\nu}={\epsilon _{\nu}+\sigma J_{\nu}\over
\kappa _{\nu}+\sigma} .
\label{eq-sourcefunction4}
\end{equation} 

For instance, via the statistical equilibrium equations,
 the source function writes for a two-level atom without
an underlying continuum, and assuming that emission and absorption
profiles $\phi$ and $\psi$ are equal (i.e., we assume complete
redistribution in frequency which makes the line source function
independent of frequency):

\begin{equation}
S({\rm line}) = {\int J_\nu \phi_\nu \, d\nu \, + \eta  \, B_\nu(T_e)
\over 1 + \eta } \ ,
\label{eq-sourcefunction5}
\end{equation}
with:

\begin{equation}
\eta  = {n_e C_{ul} \over A_{ul}}
\left[1- \exp \left(-{E_{ul} \over k T_e}\right)\right] \ ,
\label{eq-sourcefunctionter}
\end{equation}
where $B_\nu(T_e)$ is the Planck function at the electronic
temperature $T_{e}$, $n_e$ is the electronic density, and $C_{ul}$
and $C_{lu}$ being the collisional excitation and deexcitation
coefficients. {\it We see that the source function  can be separated
between a $J$-dependent and a $J$-independent component}.

Introducing now the optical depth: 

\begin{equation}
\tau_{\nu}=\int d\tau_{\nu}=\int (\kappa_{\nu}+\sigma_{\nu})dz ,
\label{eq-opacity}
\end{equation}
Eq. \ref{eq-trans1} can be written:

\begin{equation}
\mu {dI_{\nu} \over d{\tau_{\nu}}} = -I_{\nu} + S _{\nu} .
\label{eq-trans2}
\end{equation}

The formal solution of Eq. \ref{eq-trans2} is:

\begin{equation}
J(\tau_{\nu}) = {1 \over 2} \int d\mu  \int S_{\nu}(t)
e^{-(t- \tau_{\nu})/ \mu }  dt/ \mu .
\label{eq-trans3}
\end{equation}

\subsection{The ALI method}

Eqs. \ref {eq-sourcefunction4}, \ref{eq-sourcefunction5}, and
  \ref{eq-trans3}, show that the source function and the intensity are
  coupled through the transfer and the statistical equilibrium
  equations of the levels.  It requires an iterative solution, unless
  it is possible to decouple these quantities using some
  approximation, like the escape one.

The iterative process can be exceedingly long unless a fast algorithm
is used.  In the new version of Titan, the iteration is performed with the ALI method. 
 It is based on
iterative schemes using operator splitting i.e., numerical methods better
known as Jacobi's method in mathematics (see Trujillo Bueno \& Fabiani
Bendicho, 1995, for a very clear discussion of iterative methods for the
non-LTE radiation transfer problem). Such techniques were first introduced
in the field of radiation transfer by Cannon (1973). A breakthrough was
made when Olson et al. (1986) demonstrated that a fast and accurate
solution can be computed using a diagonal approximate operator; this is
what we have done in the present study (see also review of Hubeny 2001).
 We give here only a very brief account of
the method.

We can formally write, introducing the operator $\bf\large {\Lambda}$:

\begin{equation}
\int J(\tau_{\nu}) \phi_\nu d\nu ={ \bf\large {\Lambda} }[ S_{\nu}(t) ] \, .
\label{eq-transfer10}
\end{equation}
Let us now introduce the following perturbations:

\begin{eqnarray}
\cases{
{\bf\large{\Lambda}} = {\bf\large{\Lambda^*}} + ({\bf\large{\Lambda}} -
{\bf\large{\Lambda^*}}) \cr
S_\nu=S^{*}_\nu + \delta S_\nu
} \, ,
\label{eq-transfer20}
\end{eqnarray}
where $\bf\large{\Lambda^*}$ is an approximate $\bf\large{\Lambda}$
operator, and $S_{\nu}^*$ has been calculated at the previous
iteration. It is easy to derive an expression for the line source
function increment $\delta S_\nu$:

\begin{equation}
\delta S_{\nu} = [1 - {1 \over {(1+\eta)}} {\bf\large{\Lambda ^*}}]^{-1}
\left[ { {\Lambda [S_{\nu}^*] + \eta B} \over {1 + \eta}} - S_{\nu}^* \right]
\, ,
\end{equation}
where $\eta$ is defined by Eq. \ref{eq-sourcefunctionter}.

The choice of the approximate operator ${\bf\large{\Lambda ^*}}$ is
therefore critical. However, in all calculations presented hereby, we
adopted a diagonal approximate operator following Olson et
al. (1986). It is also important to mention that the formal solution
solver uses the 2nd order short characteristics method (Olson et Kunasz 1987; see
also Kunasz \& Auer 1988, Auer \& Paletou 1994); acceleration of convergence 
was also used (Ng 1974, Auer 1991).

For multi-level ions, we use in addition the so-called ``preconditioned"
equations of statistical equilibrium, following Rybicki \& Hummer (1991).

To perform this computation, one needs to know the opacities and
emissivities at each frequency as functions of $z$, hence the temperature
and the populations at each depth, so an iteration procedure is required.
Indeed the ionization and thermal equilibria  depend on the radiation 
field in a highly non-linear way. Therefore, when the optical 
thickness is large (say, for a Thomson thickness of the order of 10),
 between 30 to 100 iterations are 
necessary to ensure the stability of line fluxes and energy balance. It 
is much more rapid for 
optically thin models.

Our iterative scheme is thus as follows:

\begin{itemize}

\item For each layer, starting from the illuminated side, the 
ionization, statistical, 
and thermal balance equations are solved by iteration with $\int 
J_{\nu}{\phi}_{\nu} d\nu$, $J_{line}$ and $J_{c}$ from the previous iteration; 
the temperature, the opacity and the emissivity are computed;

\item when the back side of the cloud is reached, the transfer is solved 
with the ALI method, and $J_{\nu} (z)$ is computed for each layer of the whole slab;

\item the whole calculation is repeated until convergence;
 It is stopped when the energy balance is achieved 
for the whole slab (i.e. when the flux entering on both sides of 
the slab is equal to the flux coming out from both sides).

\end{itemize}

We stress that the choice of the grid in $z$ is critical in this 
computation, 
especially
in the case of a photoionized model, owing to the rapid decrease of temperature
and the rapid variation of the fractional ionic abundances
occuring when the X-ray photons are absorbed. The ALI method
 gives good results when the number of layers is at least equal to 
3 per decade of optical thickness. In order to fulfill this 
requirement, it is necessary to add 
several layers
at the points where the temperature and ionization degrees vary 
rapidly.
 Moreover the optical thickness of the first
 layer must be smaller
than 0.01. We use therefore a nearly symmetrical
logarithmic grid, such that the optical thickness of the first and the last 
layer is smaller than 0.01 for all the lines. 

In total, about 450 
layers are necessary to ensure a correct computation, but we have 
used also a larger number of layers with different grids in order to 
check the precision of the results. For instance, 
when the numbers of layers is doubled in the regions where the 
temperature and ionization degrees vary rapidly (which amounts at using 650 
points instead of 450) the line intensities differ at most by 
1.8$\%$ in the UV and by 0.3$\%$ in the X-ray range.

\subsection{The escape probability formalism}

A body of important literature has been devoted to the escape 
probability formalism for more than three decades
 (Hummer 1968, Hummer \& Rybicki 1970, Irons 
1978, Elitzur 1982,  Rybicki \& 
Hummer 1983\ldots). 
It is obviously beyond the scope of the present 
paper to give an account of the subject. Reviews by Rybicki, Frisch, 
Athay, Canfield et al., can be found for 
instance
in ``Methods in radiative transfer" (Kalkofen 1984).  The physical 
basis and the main 
hypothesis underlying the escape formalism are summarized 
clearly by
Hubeny (2001) in the case of a two level atom without an underlying
 continuum. It was also discussed in DAC. We will recall 
here only some ideas underlying this formalism, in order to be able to
explain in Section 5 why it does not give correct results for the
 X-ray spectrum.

\medskip 

 The essence of the escape probability method is decoupling
the source function 
(i.e. for a line the statistical equations of the levels)
and the intensity (therefore the transfer 
equation).

Let us now define the 
``Net Radiative 
Bracket", NRB, of the rate
equations, $\rho _{ul}$:
\begin{equation}
\rho _{ul}={\left\{{ n_u( A_{ul} + B_{ul}\int J_{\nu}{\psi}_{\nu}d\nu) - n_l 
(B_{lu}\int J_{\nu}{\phi}_{\nu} d\nu )}\right\} \over n_u A_{ul}}. 
\label{eq-div-flux} 
\end{equation}

Using $\rho_{ul}$, one can write Eq. \ref{eq-flux-line-integre} as:
 \begin{equation}
{dF_{line} \over dz}=  h\nu \rho  _{ul} n_u A_{ul} - 4\pi \kappa_{c} J_{line}.
\label{eq-flux-line-integrebis}
\end{equation}
Again notice that for a line with an underlying continuum, $ \rho _{ul}$ 
contains the continuum intensity.

The first term of the right side of Eq. \ref{eq-div-flux} 
represents the net cooling rate per unit 
volume (it is equal to $\ h\nu \ 
(n_u\ n_{\rm e} C_{ul}\ - n_l\ n_{\rm 
e}C_{lu})$ {\it only} for a two-level atom), while the second member
is the heating rate by continuum absorption of line photons.  
This equation shows that when $\kappa_{c}$ or $ J_{line}$ is small,
 the line flux is equal to that of
an optically thin 
medium, multiplied by ${\rho}_{ul}$. 

The frequency-dependent 
transfer equations are now
transformed into frequency integrated ones, which can be solved consistently 
with the statistical rate equations and the energy equilibrium equations.
The challenge is to replace ${\rho}_{ul}$ by an expression 
independent of the 
local line radiation field. 
The escape probability formalism performs this task by
 {\it identifying with $\rho$
 the probability 
$P_{e}$ 
of a photon
 emitted at 
an optical depth $\tau$ from the surface to 
escape  in a single flight}.  It is used generally in the case of a 
line, but it can be used also for the continuum, like in XSTAR.

The escape probability of a photon of frequency $\nu$ emitted at
 an optical thickness $\tau_{\nu}$ is equal to:
\begin{equation}
P_{e}(\tau_{\nu})= e^{-\tau_{\nu}/\mu},
\label{eq-escape0} 
\end{equation}
and the angle averaged escape probability:
\begin{equation}
P_{e}(\tau_{\nu})= {\cal{E}}_2(\tau_{\nu}) ,
\label{eq-escape00} 
\end{equation}
where ${\cal{E}}_2$ is the second order integro-exponential. 
 In the case of a 
slab of 
finite thickness, radiation is emitted by both sides, 
and the total
escape probability becomes: 
\begin{equation}
P_{tot}(\tau_{\nu})={ P_{line}(\tau_{\nu}) + P_{line}(T_{tot}-\tau_{\nu})\over 2}
\label{eq-escapetot} 
\end{equation}
where $T_{tot}$ is the total optical thickness of the slab at the
frequency $\nu$.

\medskip

Let us consider a line photon with an emission profile $\phi_{\nu}$ 
and no underlying continuum. Integrating over the frequencies, one gets:
\begin{equation}
P_{line}(\tau)=  \int_0^\infty {\cal{E}}_2[\tau{\phi_{\nu}}] 
{\phi}_{\nu}\ d\nu
\label{eq-escape} 
\end{equation}
and we 
call $P_{escl}$ the escape probability from both sides (to be 
homogeneous with some authors). Then, the right member of Eq. 
 \ref{eq-flux-line-integrebis} is reduced to the first term, and the 
 line fluxes can be computed easily provided that a correct expression 
 for $P_{line}$ is used. Moreover one has to take into account the escape probability
  only in the rate equations of the levels and in the line flux, as there is 
  no 
additional heating 
or ionization term due to the line in the energy of ionization 
equilibrium.

For complete redistribution in a Voigt profile, one finds from Eq. 
\ref{eq-escape} that $P_{line}$
 can be written approximately 
(cf. Collin-Souffrin et al. 1981): 
\begin{eqnarray}
&&  P_{line}(\tau_0) ={1\over 1+2\tau_0\sqrt{\pi 
ln(\tau_0+1)}},\ \  \tau_0 \le 1/a
\\ \nonumber
&&  P_{line}(\tau_0) ={2\over
3}\sqrt{{a\over\tau_0\sqrt{\pi}}}, \ \ \tau_0 > 1/a
\label{eq-esc-1}
\end{eqnarray}
where $\tau_0$ is the optical thickness at the line center and $a$ is 
the usual damping constant. The first expression corresponds to
the Doppler core, and the second to the Lorentz wings of the 
Voigt profile, which become important only for large values of $\tau_0$. 
Using this expression in Eq. \ref{eq-escapetot}, one gets the total escape 
probability replacing $\rho_{ul}$ in Eq. 
\ref{eq-flux-line-integrebis}.

Redistribution in 
frequency may be incomplete for intense lines, 
whose wings are 
optically thick, and/or for strongly interlocked lines. {\it This is impossible 
to account for without a coarse approximation with the escape 
probability formalism}.  
Note however that only very thick 
lines are dominated by the Lorentz wings (basically if $\tau_0 \gg 1/a$). 

Summarizing the assumptions, the identification of 
$P_{line}$ with $\rho _{ul}$ implies a
homogeneous medium (it is a ``global" approximation, which is used 
``locally"), that the absorption and emission profiles 
 $\phi_{\nu}$ and $\psi_{\nu}$ are identical (i.e. if there is 
complete redistribution of frequencies), 
 and no destruction mechanisms of line photons (no 
 second term in the right member of Eq. 
 \ref{eq-flux-line-integrebis}). Note in particular {\it a crucial 
 difference between the $\rho$ and $P_{line}$, namely that the first 
 quantity
 can become negative, 
 while the latter is always positive}. Hubeny (2001) showed that {\it even in 
 this oversimplified case, the result of the escape probability computation 
 for the source function and the emergent 
 profile and intensity  differ significantly from that of the full 
 transfer}, because the escape probability approximation
  fails in the outer layers where the line core 
 is formed and the source 
 function varies strongly (his Figure 1). 

\medskip 
  
Moreover the emission in AGN and in X-ray binaries is
 very far from this simplified case of a two-level atom 
and a
homogeneous medium. In a highly ionized medium with a
high column density, 
 an intense diffuse continuum underlies the X-ray lines, which on 
 the other 
 hand are reabsorbed in photoionizing less ionized species.
  The escape probability should thus take 
  into account both destruction of line photons and radiative 
  excitation by continuum photons. As we will see in the next 
  section, there is no consensus on the way this should be done.
 
\section{Different approximations used for the escape 
probability.}

 Several 
approximations are used in the literature 
to account for 
these effects.
 The reader is invited to refer to original papers for more detailed 
explanations. 

\subsection{In the absence of an underlying continuum}

 XSTAR (and consequently 
Nayakshin's code) 
assumes 
complete redistribution in a Doppler core for all lines 
(cf. Kallman and Bautista 2001),
 with $P_{line}$ given
from Hollenbach \& McKee (1979):

\begin{eqnarray}
P_{line}(\tau)&=&{1\over \tau\sqrt{\pi}(1.2+{\sqrt{log(\tau)}\over 1+10^{-5}
\tau})},\ \ (\tau \ge 1)
\\
\nonumber
P_{line}(\tau)&=&{1-exp(-2\tau)\over 2\tau},\ \ (\tau \le 1). 
\label{eq-esc-2} 
\end{eqnarray}
where $\tau$ is here the mean optical thickness in the line,
 equal to $\sqrt{\pi}\tau_0$.

Other approximations are adopted. For instance, in Cloudy 95, incomplete 
redistribution is mimicked by complete 
redistribution in a Doppler profile for resonance lines 
such as CIV, and L$\alpha$ of helium and 
hydrogen (which are intense in the Broad Line Region of AGN for which 
Cloudy is appropriate, and weak in 
regions emitting the UV-X continuum like those concerned by this paper). 
In some approximations $P_{line}$ is derived from Bonihla
 et al. (1979) and 
Hummer and Kunasz (1980), and is (see Rees et al. 1989):

\begin{equation}
P_{line}(\tau_0) = {1\over 1+ b(\tau_0)\tau_0}, 
\label{eq-esc-3}
\end{equation}
where $b(\tau_0)$ is a factor of the order of unity, differing in the Lorentz 
wing and in the Doppler core.
The other lines are treated by complete redistribution in Voigt profile, 
 with an expression similar to  Eq. \ref{eq-esc-1}. 

Finally Ross et al. (1978) uses

\begin{equation}
P_{line}(\tau_0) = {1\over 4\tau_0(ln\tau_0/\sqrt{\pi})^{1/2}}. 
\label{eq-esc-4}
\end{equation}

Since our reference model in the following computations is a slab of finite 
thickness, we will use mainly Eq. \ref{eq-escapetot} which takes into account the emission 
from both sides.   
Except in one case where we will use Eq. \ref{eq-esc-3}, this equation will be coupled 
with Eq. \ref{eq-esc-1},  
coherent with our 
solution of the transfer equation assuming complete redistribution in a 
Voigt profile. Note that the X-ray lines in which we are interested here
 have a relatively small optical thickness 
($\tau_0\le 10^{5}$, i.e. $\le 1/a$), so the difference between a
Voigt and a Doppler profile is not important.

\subsection{In the presence of destruction processes}

As mentioned above, things are complicated by the fact that line photons 
are 
destroyed by several 
processes, in particular in photoionizing less ionized species. 
One must thus take into account these
photoionizations and their contribution to the 
energy balance. 

Continuum opacity is taken into account in two different 
ways. In some codes (XSTAR, Cloudy), $P_{line}$ is replaced by
$F(P_{line})P_{line}$, where $F$ is an operator given by Hummer (1968) 
 accounting for destruction by continuum absorption in one line scattering: 

\begin{equation}
F(X)=\int_{-\infty}^\infty {\phi (x)\over 
X+\phi (x)} dx
\label{eq-FX}
\end{equation}
where 
$x=\delta \nu /\delta \nu _D$, and $\delta \nu _D$ is the Doppler width.

In ION (Netzer et al. 1985), as well as in old versions of Cloudy 
(cf. Ferland \& Rees, 
1988, and Rees et al. 1989), $A_{ul}$ is
replaced  in the level equations
 by $A_{ul}\times P_{esc}$, with:
\begin{equation}
P_{esc} = P_{line}(\tau_0+\tau_c) \times {\kappa_0\over 
\kappa_0+\kappa_c} +  {\kappa_c\over 
\kappa_0+\kappa_c},
\label{eq-esc-7}
\end{equation}
 where $\kappa_0$ is the absorption coefficient at the line center. In the 
 most recent version of Cloudy, the operator $F$ is also introduced. 
 
\subsection{Local and non-local processes}

It was shown a long time ago (Hummer and Rybicki 1970) that in the 
absence of a destruction mechanism, a line 
photon is scattered many times inside the line core, 
staying close to its point of emission. 
And when finally the photon is driven in the line wing by scattering, 
it escapes from the 
medium in one 
single long flight.

 This is actually the foundation on which all the codes are 
built using the escape probability 
approximation. The medium is divided in a number of small layers 
(generally a few hundred). Once line photons have been emitted 
in a given layer, they escape from this layer with a probability 
$P_{esc}$ (which is computed for the {\it whole} 
medium), and then they {\it 
lose their identity of line 
photons to 
become continuum photons}, which are no subject to any line 
processes. 

Moreover, destruction 
mechanisms such as photoionizations or Compton process,
 can take place both during the 
diffusion of the line photons close to their point of emission, and on their 
way towards the surface
 and towards the back of the 
cloud.
It is thus
 necessary to differentiate between ``local" and ``non-local" 
 processes. 
 
\subsubsection{Local processes}

\medskip

\noindent{\bf Rate equations}

\medskip

Ko \& Kallman (1994) adopt the following scheme to solve the line transfer in 
the irradiated atmosphere of an accretion disc (the same approximation was 
probably used in the previous version of XSTAR). If a line photon does not 
escape,  it has a probability 
$(\kappa_c+\sigma)/(\kappa_l+\kappa_c+\sigma)$ of being destroyed 
locally by continuum 
absorption, where $\kappa_l$ is the {\it mean } absorption line 
coefficient
=$\sqrt{\pi}\kappa_0$, and 
$\kappa_c$ is the continuum absorption coefficient
 at the line 
frequency, due to photoionization and free-free processes. 
In the statistical level equations, $A_{ul}$ is thus
replaced by $A_{ul}\times P_{esc}$, with:
\begin{equation}
P_{esc}= P_{line}(\tau) +  (1-P_{line}(\tau)) {\kappa_c+\sigma\over 
\kappa_l+\kappa_c+\sigma},
\label{eq-esc-5}
\end{equation}
where $\tau$ is the mean line opacity. 

In the last version of XSTAR the approximation is slightly different 
(cf.  Kallman \& Bautista 2001).
 $\kappa_c/\kappa_l$ 
and $\sigma/\kappa_l$ are replaced respectively by $\kappa_c/\kappa_l\times 
F(\kappa_c/\kappa_l)$ and by $\sigma/\kappa_l\times
F(\sigma/\kappa_l)$.
 Then the expression used 
in the statistical level equations is:

\begin{equation}
P_{esc}= min[1, P_{line}(\tau) \times (1+{\kappa_c\over 
\kappa_l}F({\kappa_c\over \kappa_l}))].
\label{eq-esc-6}
\end{equation}

In Ross et al. (1978) the escape probability in the statistical level 
equations is written:

\begin{equation}
P_{esc} = {P_{line}(\tau_0)\over P_{line}(\tau_0)+(\kappa_c/\kappa_0)\times
 F(\kappa_c/\kappa_0)}
\times e^{(-\sqrt{3}\tau_c)}. 
\label{eq-esc-8}
\end{equation}

\medskip

\noindent{\bf Local energy and ionization balance equations}

\medskip

 In some cases, all photoionizations by line photons are assumed to take place 
locally (``on the spot"), and consequently also the corresponding 
 energy gains. 
This is the case of Cloudy and of 
Ko \& Kallman computations (1994). This treatment 
is not appropriate if the line photons are reabsorbed in a 
location where the physical 
conditions are different from those of the emission point, as 
in Thomson thick photoionized media, which are generally
very inhomogeneous in temperature and in ionization state.

Ionization by line photons corresponds to a cooling term in the local energy
 balance. It can 
 be expressed as:
\begin{equation}
\Lambda_{line} = {n_uA_{ul}h\nu\over n_{e}n_{H}} \times \beta,
\label{eq-beta}
\end{equation}
where $\beta$  includes or not continuum absorption.
 Ko and Kallman (1994) use a term $\beta= P^{'}_{esc}$, with:

\begin{equation}
P^{'}_{esc}= P_{line}(\tau) +  (1-P_{line}(\tau)) {\sigma\over 
\kappa_l+\kappa_c+\sigma},
\label{eq-esc-9}
\end{equation}
 This expression takes into account 
the fact that a line photon that does not escape directly has a probability of 
being removed from the line core by Compton scattering but still contributes 
to the line intensity. Since  continuum 
absorption is taken into account locally, the emergent line intensity on 
one side
(frequency integrated) is obtained with 
a simple integration
\begin{equation}
4\pi J_{\rm line}=\int{n_uA_{ul}\times P^{'}_{esc}\ dz},
\label{eq-esc-10bis}
\end{equation}
where the integration is performed between the point of emission and the 
surface.

This treatment is different from that of XSTAR, where $\beta= P_{line}(\tau_l)$,
 i.e. continuum absorption is not taken into 
 account locally in the energy balance and in the line emissivity (as 
 far as we understand). 

A hybrid treatment was adopted by Ferland and Rees (1988) as $A_{ul}$ 
is replaced  in the energy
 balance equation by:
\begin{equation}
A_{ul}\times 
P_{line}(\tau_0+\tau_c),
\label{eq-esc-11}
\end{equation}
which takes into account
 continuum absorption locally in the thermal balance, but 
at the same time the line photons are attenuated by the exponential 
factor (only towards the illuminated side). 
They take into account a local ionization rate by line photons:
\begin{eqnarray}
Ion&=& n_uA_{ul}[ P_{line}(\tau_0+\tau_c) \times {\kappa_0\over 
\kappa_0+\kappa_c} 
\\
\nonumber
&+& {\kappa_c\over 
\kappa_0+\kappa_c} - P_{line}(\tau_0+\tau_c)\times {\rm exp}(-\tau_c)]
\ \ {\rm cm^{-3} s^{-1}}.
\label{eq-ion-rate2}
\end{eqnarray}

\subsubsection{Non local processes, emerging fluxes}

As already mentioned, in the escape approximation the fraction of line photons 
that does not escape 
 directly from their point of emission is injected in the continuum and 
 transferred like continuum photons. For instance in XSTAR 
these 
 photons are exponentially 
attenuated by non-local continuum absorption:
\begin{equation}
4\pi J_{\rm line}=\int{n_uA_{ul}\times P'_{esc}\times  e^{-\tau_c}dz}
\label{eq-esc-10}
\end{equation}
where $\tau_c$ is the optical thickness between the point of 
emission and the surface. 
These line photons are taken into 
account in the ionization and gain processes far 
from the emission point. This is performed also in Nayakshin's and 
Ross-Ballantyne's 
codes. 

Of course this method allows us to take into account
heating and ionization due to the line photons far from the emission point, 
and to provide better emerging line fluxes than the pure local 
treatment. In spite of that, it is not
fully self-consistent.
 
Indeed, to perform calculations, the slab is divided into
 a few hundred layers, in order to get a good mesh for 
the optical thicknesses at all important frequencies of the continuum, in 
particular 
at the ionization edges. This division is made independently of the 
lines: a given layer is optically thick for some lines, optically thin for others.
In the escape approximation, whatever the thickness of the layer, it is decided
 that if a line is emitted there, a fraction $P_{esc}$ escapes the layer. 

First this fraction is
 computed according to the {\it total} optical thickness of the 
lines, which depends on the conditions out of the layer, in the whole 
medium. As we have also 
seen, the functional dependence of $P_{esc}$ on $\tau$ is hampered by 
many uncertainties and approximations.

Second, one has 
to decide if a fraction of the line is reabsorbed by a continuum process inside the layer 
itself, and we have seen that the approximations differ on 
this point, for instance on the way Thomson diffusions are taken 
into account inside the layer or not.

Finally, since line photons are treated as continuum photons out
 of the emission layer, 
it means that they will not be able to reexcite the line. In a sense, one can 
say that {\it any escape approximation involves assuming somewhat arbitrarily 
the fraction of line photons that are absorbed ``on the spot" and 
non-locally}. 

For all these reasons, the escape probability approximations do {\it not 
allow us to achieve 
 a total energy balance}, i.e. the total 
energy absorbed by the medium (from the incident spectrum) will not be 
exactly equal 
to the total
energy emitted by the medium,  {\it in spite of the fact that the local energy balance
can be achieved 
with a very high precision}. This was already stressed in DAC, and it 
will be illustrated in the next section.

\begin{figure}
\begin{center}
\psfig{figure=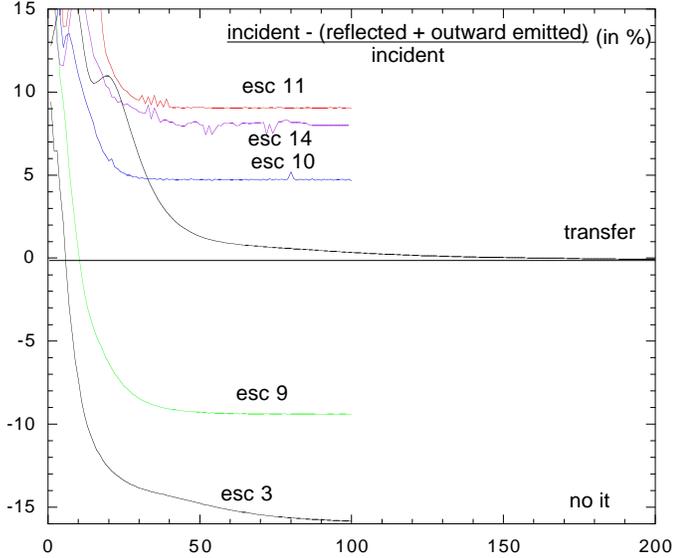,width=9cm}
\caption{This figure shows the total net energy balance for the 
reference model, i.e. the difference 
between the total incident flux and the sum of the reflected and 
outward fluxes, referred to the incident flux, as a function of the number of 
iterations, with the different approximations and with the full transfer 
treatment. We do not show Escape 
12, which converges towards +47$\%$, and Escape 13, which gives 
almost the same result as Escape 14. It shows that a large number of 
iterations are necessary for the convergence of Titan, which reaches 
then total energy balance, while the escape probability approximations 
converge more rapidly, but do not reach energy balance. }
\label{fig-bilan-ite}
\end{center}
\end{figure}

\section{Escape probability versus transfer models}

\subsection{The models and the approximations used}

        Our reference model is a plane-parallel slab of constant density $n=10^{12}$ 
 cm$^{-3}$ and total hydrogen
 column density $5\times 10^{24}$ 
 cm$^{-2}$ (Thomson thickness $\tau_{es}= 4$), irradiated on one side 
by a semi-isotropic incident continuum with a spectral distribution
 $F_{\nu}\propto \nu^{-1}$  extending from 0.1 eV to 100 keV.
The ionization 
parameter
$\xi$ at the surface of the irradiated slab,
defined as
$\xi = 4 \pi\ F/ n$,
where $F$ is the integrated incident flux and $n$ is the number density,  
is equal to $10^3$. The equilibrium temperature in such a slab (see 
below) varies between 10$^{5}$ to 10$^{6}$ K, so such
 a model is typical for the region emitting  the ``Big Blue Bump" in AGN, 
and for the ``irradiated 
skin" of an accretion disc giving rise to the reflection spectrum 
observed in the X-ray range (Nayakshin et al. 2000). We also 
considered a thinner slab (column density $10^{24}$ 
 cm$^{-2}$, i.e. $\tau_{es}\sim 1$) to see whether the escape 
 approximation is more valid in this case.
 
 Finally it is worth noticing that, keeping the same ionization 
 parameter, {\it we did not succeed in running any escape 
 approximation with a slab  thicker than $\tau_{es}=5$} (the energy 
 balance does not converge). 
 
        In order to compare the structure and the emission spectrum computed with 
the escape probability approximation and with a real line 
transfer, we performed computations using, on one side the recently updated version of 
the code Titan (cf. Coup\'e 
2002 for the atomic data) coupled with the NOAR code as described in DAC, 
plus ALI, and on the other side the same
 version (same atomic data, same 
transfer treatment of the continuum), but with the escape 
probabilities replacing the 
corresponding terms in the statistical and ionization equilibrium equations, 
in the energy 
balance and in the emitted line fluxes. 

We have constructed a set of approximations. We give here neither 
the description nor the results 
corresponding to 
all our computations, but only those of the most representative ones, which 
follow
 as closely as possible  the treatment  
of Ko \& Kallman (1994), of Kallman \& Bautista (2001), i.e. XSTAR,
 and of Ross-Ballantyne's and 
Nayakshin's codes.

\begin{figure}
\begin{center}
\psfig{figure=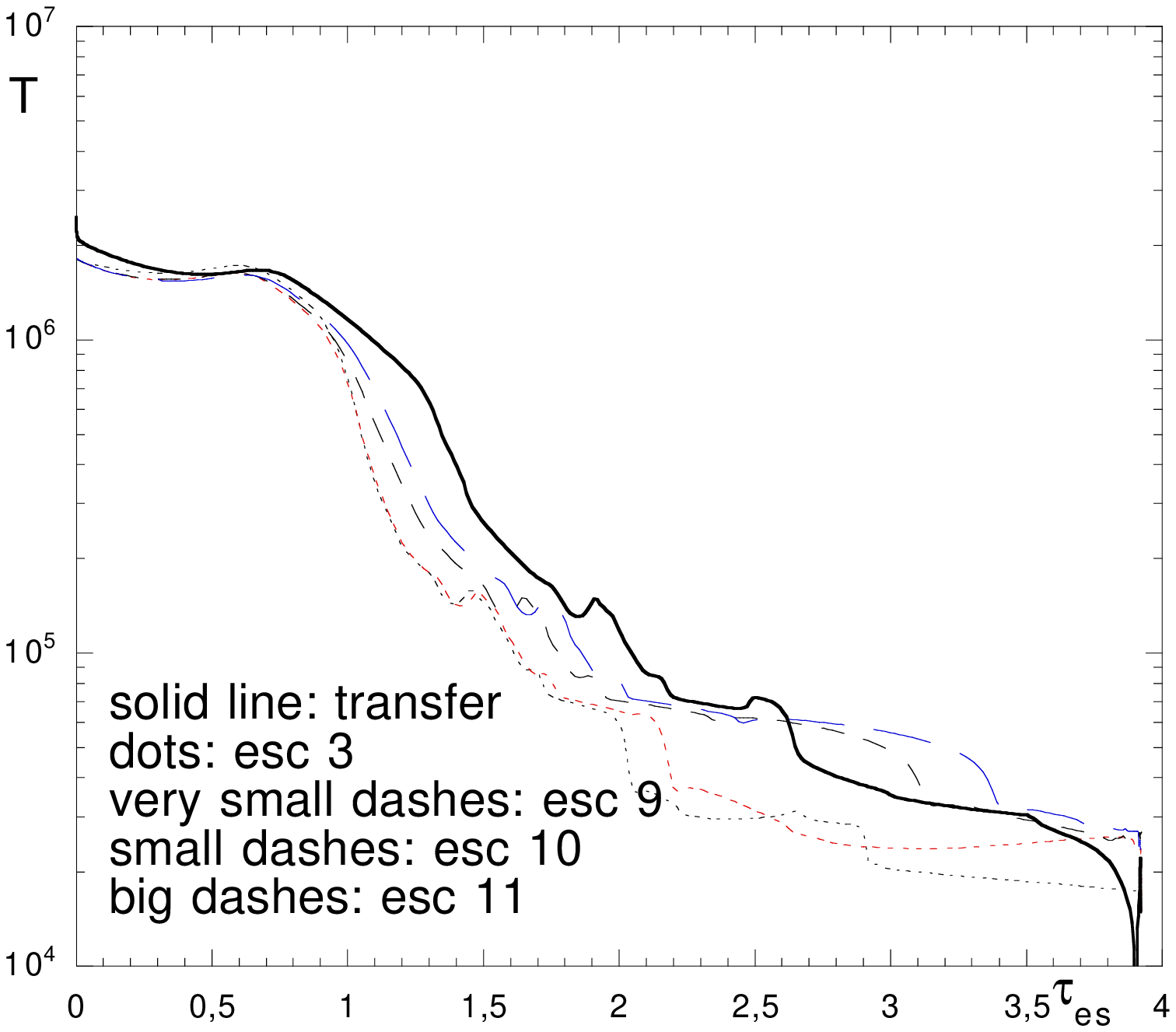,width=9cm}
\psfig{figure=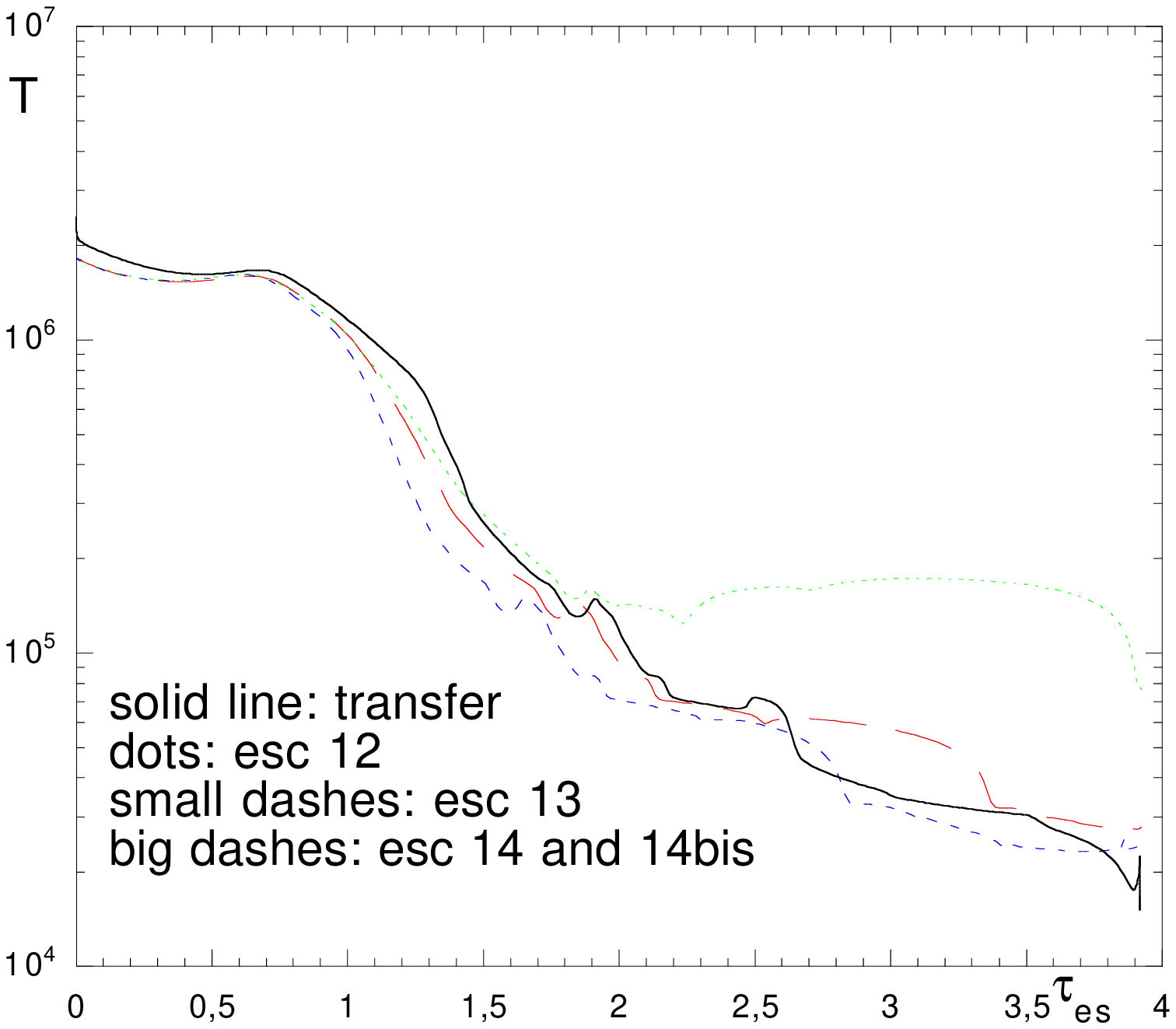,width=9cm}
\caption{Temperature versus $\tau_{es}$ for the reference model, with
 the different approximations 
and with the full transfer computation. The 
escape approximations give almost the same result as the transfer in 
the hot layers, but not in the colder ones.}
\label{fig-T}
\end{center}
\end{figure}

\begin{figure}
\begin{center}
\psfig{figure=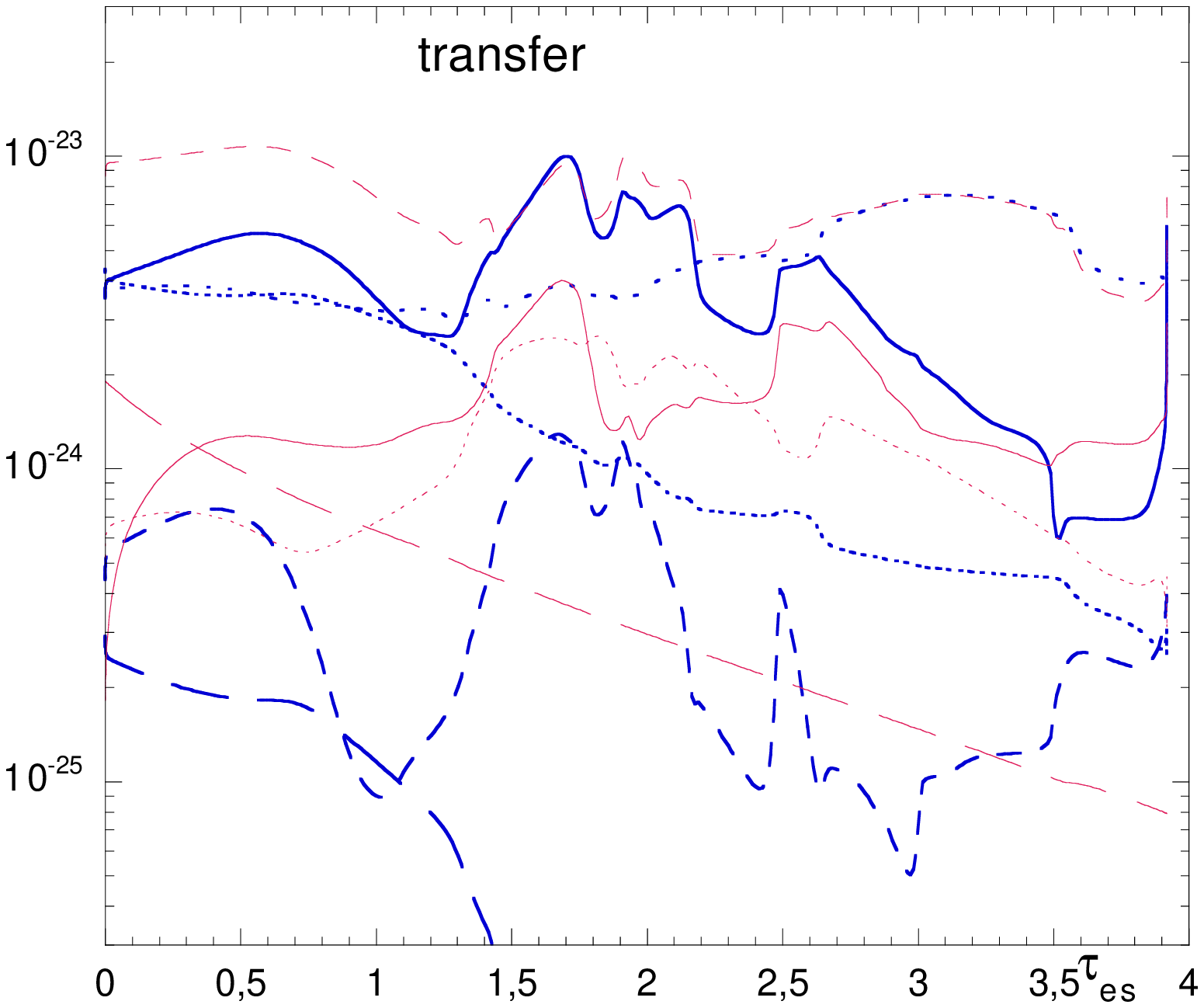,width=8cm}
\psfig{figure=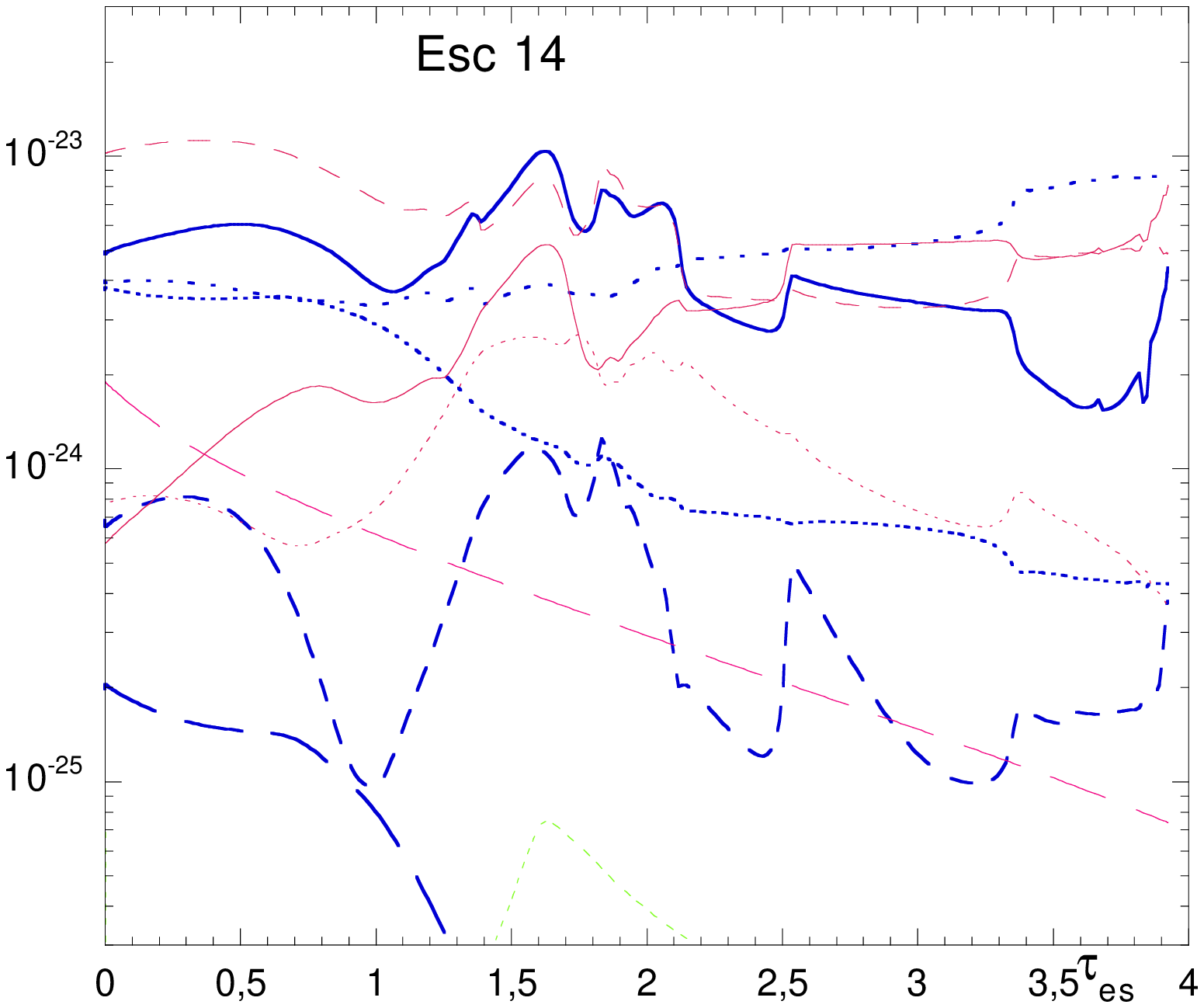,width=8cm}
\caption{Heating and 
cooling rates due to the different processes, in ergs cm$^{+3}$ 
s$^{-1}$,  versus $\tau_{es}$, for the reference model, with the full transfer computation and 
with Escape 14 approximation. 
Thick lines correspond to cooling and thin lines to heating: solid lines: 
 heating by 
photoionizations by line photons
 and net line cooling; large-dashed lines: Compton heating and cooling; 
 small-dashed 
 lines: heating by photoionizations due to continuum photons, cooling
 by radiative recombinations onto 
excited levels; 
very small dashed line: cooling by recombinations onto ground levels; dotted 
lines: free-free heating and cooling. We see that
heating by line photons is much more important for $\tau_{es}\ge 2$ 
with the 
escape approximation than with the transfer treatment.}
\label{fig-PG}
\end{center}
\end{figure}

\begin{itemize}

\item {\bf Escape 3}: line escape probability computed with 
 Eqs. \ref{eq-escapetot} and \ref{eq-esc-1},
energy 
balance with Eq. \ref{eq-esc-9}, line fluxes with Eq. \ref{eq-esc-10bis}, 
level populations with  
Eq. \ref{eq-esc-5}. 
This is a treatment similar to that of Ko \& Kallman (1994), except that
 they consider a semi-infinite atmosphere, while we are considering a 
finite slab with escape from both sides. Also, we added a term  of
{\it local} ionization due to line photons:  
\begin{equation}
Ion= n_uA_{ul}\ P_{line}\times {\kappa_c\over 
\kappa_0+\kappa_c} 
\label{eq-ion-KK}
\end{equation}
and the corresponding energy gain term. Thus it is a pure 
local treatment.

\item {\bf Escape 8}: line escape probability computed with 
 Eqs. \ref{eq-escapetot} and \ref{eq-esc-1}, $P_{esc}$ computed with the 
 Hummer operator, 
energy 
balance with only $P_{line}$ and line fluxes with Eq. \ref{eq-esc-10},
level populations with  Eqs. \ref{eq-FX} and \ref{eq-esc-6}.  
This treatment is similar to that of  Kallman \& Bautista (2001) 
with the exponential attenuation of the line photons towards the 
surface, and their 
non-local use for ionizations and gains of energy. The only 
difference is in 
the expression of the line escape probability 
(Eqs. \ref{eq-escapetot} and 23
instead of Eq. 25).

\item {\bf Escape 9}: same as Escape 8, except that the line photons 
are taken into account in photo-absorption in both directions.

\item {\bf Escape 10}: same as Escape 8, except that the ionization and gain 
equations, $P_{esc}$ is replaced by $P^{'}_{esc}$ of Kallman \& 
Bautista (2001).

\item {\bf Escape 11}: same as Escape 9,  except that the ionization and gain 
equations, $P_{esc}$ is replaced by $P^{'}esc$ of Kallman \& 
Bautista (2001).

\item {\bf Escape 12}: same as Escape 11, but using Eq. \ref{eq-esc-2} 
as Kallman \& Bautista (2001) for the escape probability.

\item {\bf Escape 13}: same as Escape 11, but using 
$\tau_{e}=\sqrt{3\tau_{abs}(\tau_{abs}+\tau_{es})}$ instead of 
$\sqrt{3}\tau_{abs}$ in the attenuation of the continuum photons. 
Note that the term $\sqrt{3}$ is present here to mimic our 
``semi-isotropic" treatment. Indeed Titan solves the 
transfer for a semi-isotropic radiation field, and in particular it 
considers an isotropic
incident radiation, instead of a monodirectional intensity perpendicular 
to the surface as the other codes.

\item {\bf Escape 14}: same as Escape 13, but with $\tau_{0}$ replaced 
by  $\sqrt{3}\tau_{0}$ in $P_{line}$ (Eqs. \ref{eq-escapetot} 
and 23),
to better fit the transfer treatment.

\end{itemize}

 We have also tried an 
approximation identical to Escape 14 (Escape 14bis), except that the 
differential line fluxes are multiplied by a factor of two, which should 
be present to take into account the semi-isotropy of the radiation 
field. This approximation is therefore more rigorous,
 we will see 
however that it leads to stronger 
lines than the other approximations, much stronger than 
 the full transfer 
computation. To clarify what we have done, we give in Appendix B more details about
the full set of equations used in the last approximation, which we 
consider as being the best.

In summary, Escape 3 is close to the treatment of Ko \& Kallman (1994), 
and is the only one to take into account for local escape by Compton diffusions in the 
statistical level equations, 
Escapes 8 and 10, 12, are close to that of Kallman \& Bautista (2001) 
and of XSTAR, and
Escape 11 and the following ones can be compared to the treatment
of Ross-Ballantyne's and Nayakshin's codes. Since Escape 8 differs from 
Escape 10 exactly like Escape 9 differs from Escape 11, we do not show the 
results corresponding to this approximation.

\subsection{Results: the structure}

\begin{table}
\begin{center}
\psfig{figure=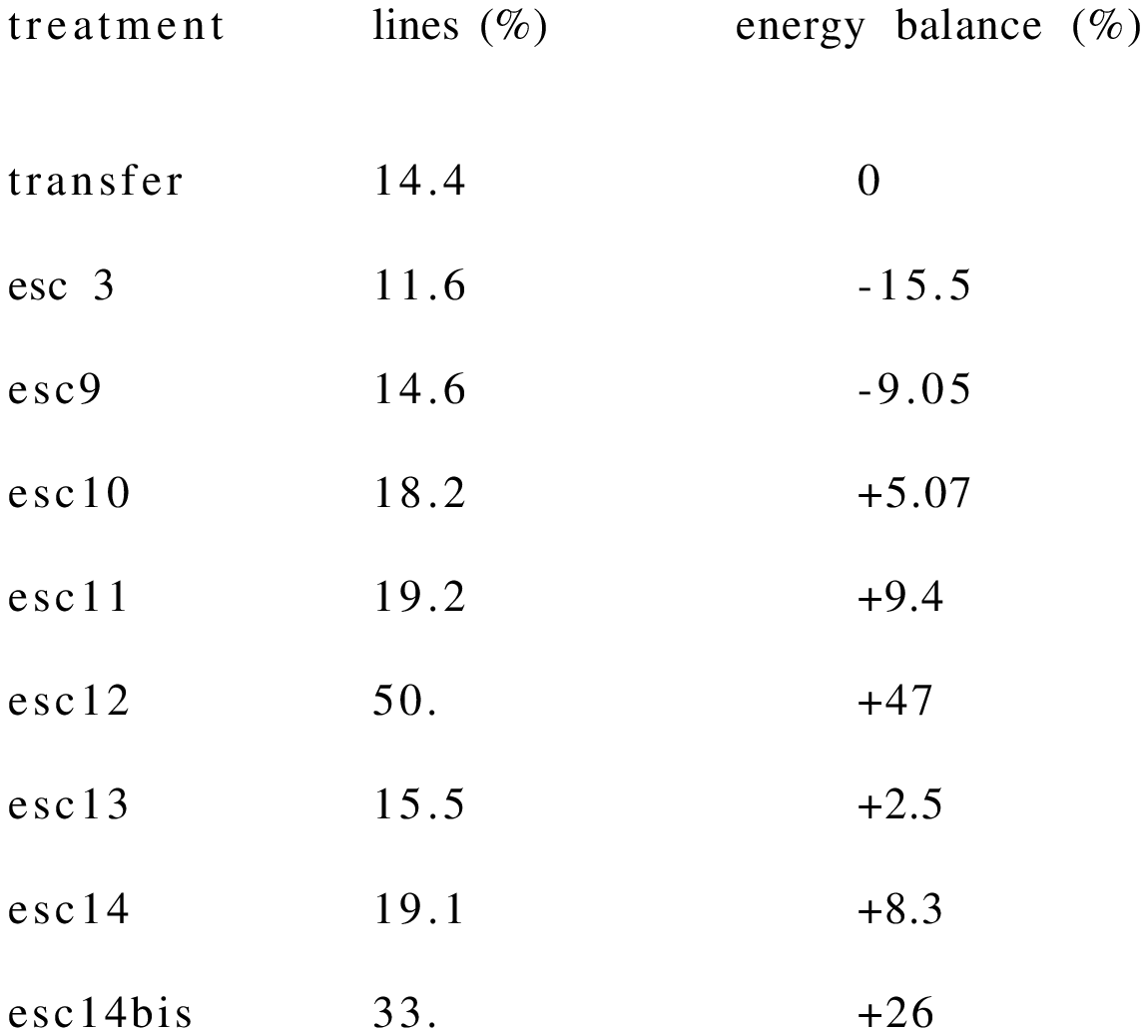,width=9cm}
\caption{Energy balance after convergence, for the reference model, 
with the different 
approximations and with the full transfer.}
\label{table-bilan}
\end{center}
\end{table}

The computations performed with an escape 
probability approximation converge more rapidly
than those 
 with the full transfer. It 
was expected, since the lines require the convergence of only
 $\tau_{0}$ and $T_{tot}-\tau_{0}$ with the escape probability 
approximations. But this is actually only a 
 {\it ``pseudo-convergence"}, as 
 the {\it total} energy balance, i.e. the difference between the 
incident flux and the emitted plus transmitted one, has a value different from zero.
It is due to the reasons 
 mentioned in the previous section.  Fig. 
 \ref{fig-bilan-ite} displays this quantity 
 referred to the incident flux (in percentage), 
 as a function of the number of iterations. We see that 
 the various approximations ``converge" towards 
different values.
From this point of view, the best approximation 
is Escape 10 (the balance is achieved within 5$\%$). 
Notice that Escape 3 is not completely ``converged" after 100 iterations.

This absence of energy balance is linked with
the discrepancy between 
the spectra, and in particular between the line fluxes,
 obtained in the case of the escape approximations and of the full transfer
 computation. To illustrate this fact, Table \ref{table-bilan} gives the number of the 
 approximation  
 (column 1), the percentages 
 of the line flux  (reflected and outward)
 with respect to the incident flux (column 2), and the mismatch in energy balance 
 in percents (column 3).
   With the transfer treatment, the spectral lines represent only 
 14$\%$ of the total 
 emission, which we assume to be the correct 
 value. We notice that the importance of the spectral lines is larger when the 
 mismatch is 
 larger and positive, indicating that it is partly due to the spectral lines. 
 The other part is due to the continuum emitted outward.

The mismatch in energy balance has a relatively small 
influence on  the structure in temperature $T$ in the hot layers, 
but not in the 
deepest layers (Fig. 
\ref{fig-T}). 
This is because the 
energy balance is dominated by continuum processes,
 which are treated in the same way
in the escape approximations and in the full transfer treatment. In the deepest layers, 
bound-bound transitions dominate
both the cooling and the heating, owing to the 
smaller temperature, and the temperature equilibrium is therefore more sensitive 
to the approximation used. 

\begin{figure}
\begin{center}
\psfig{figure=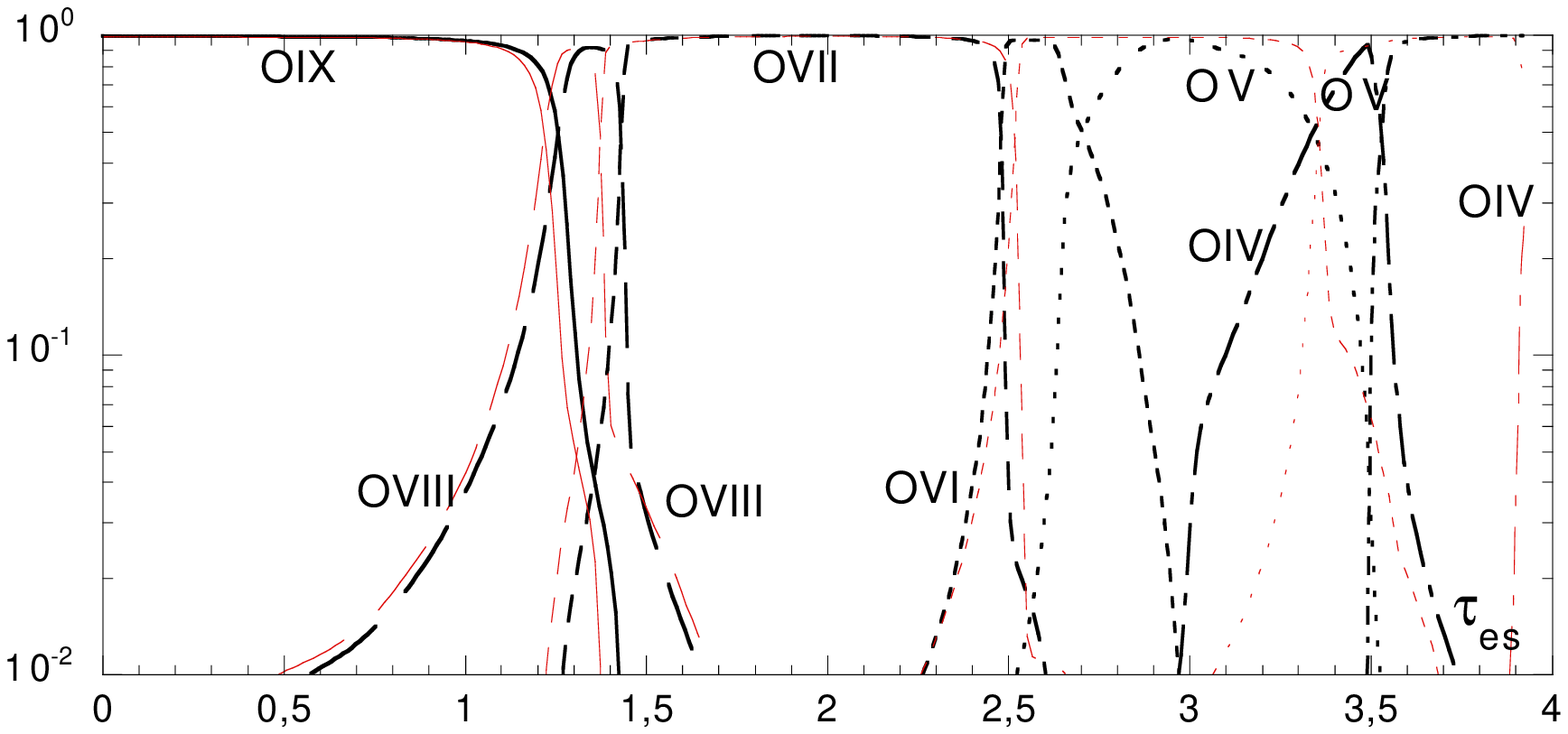,width=9cm}
\psfig{figure=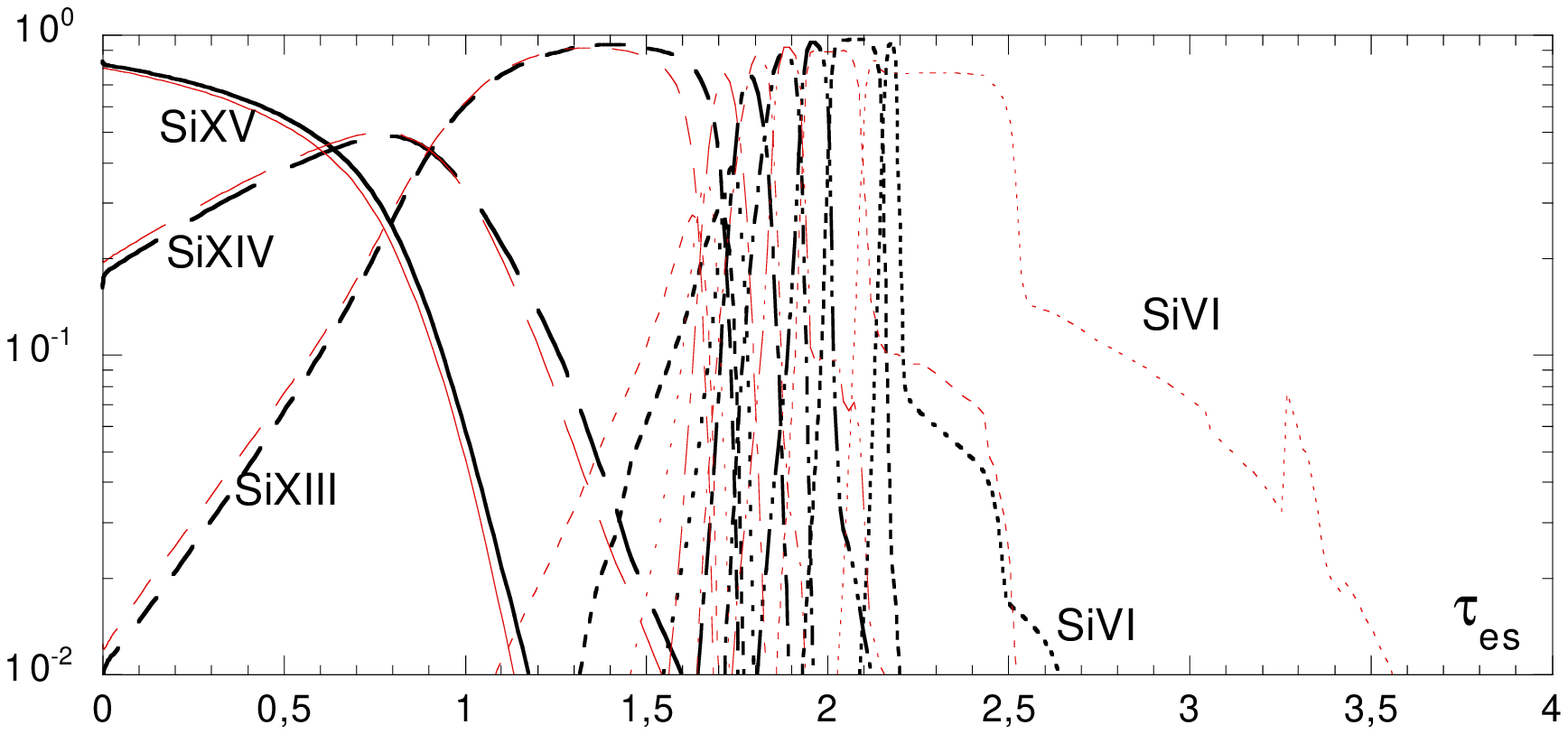,width=9cm}
\caption{Fractional abundances of oxygen and silicium as functions of the optical 
thickness,  for the reference model,
with the full transfer computation (thick lines) and with Escape 14 
 approximation (thin lines). Ionization is strongly overestimated with the 
 escape approximation for $\tau_{es}\ge 2$.}
\label{fig-ion-O}
\end{center}
\end{figure}

This can be seen in  Fig. \ref{fig-PG} which
displays the local energy gains and losses due to the different 
processes for the full transfer computation and for Escape 14 
approximation.  One can also check that
 the ``local energy 
balance" is achieved with a very good precision  in all 
computations (actually better than 10$^{-4}$).

Though the losses and gains are quite similar in the transfer and in 
the escape treatment near the surface, one notices that in the region
 where $T$ is small (i.e. for $\tau_{es}\ge 2$), heating by photoionizations due to 
 line 
photons, and line cooling, are important in the energy balance with the 
escape treatment. In particular the gains are dominated by photoionizations 
due to 
line photons, which are negligible in the transfer 
treatment.  We will see later that X-ray 
line emissivities are overestimated in the escape treatment. It
leads to the overestimation of photoionizations by line photons. 
Note in passing that lines of lithium-like
 ions have a role in the  
interplay between heating and cooling, and in the ionization equilibrium.
 First the medium 
is copiously cooled by lithium-like ions of elements 
like CNO, whose first excited levels have low potential energies (for 
this reason it is important to take into account  carefully  the excited 
levels of these ions). Second,
 owing to their relatively small ionization potential (for 
instance the ionization 
potential of OVI is 130 eV), they can  easily be collisionally ionized 
at $T\sim 10^5$K.

As examples Fig. \ref{fig-ion-O} displays the fractional abundances of oxygen and 
silicium as functions of the depth in the layer. Again the ionization equilibrium is
 little dependent on the 
approximation used near the surface, with the transfer and the escape 
treatments, but it is quite different in the deepest layers: the 
ionization state is overestimated in the escape treatment, and this 
is also linked to the overestimation of 
photoionization by line photons.

\begin{figure}
\begin{center}
\psfig{figure=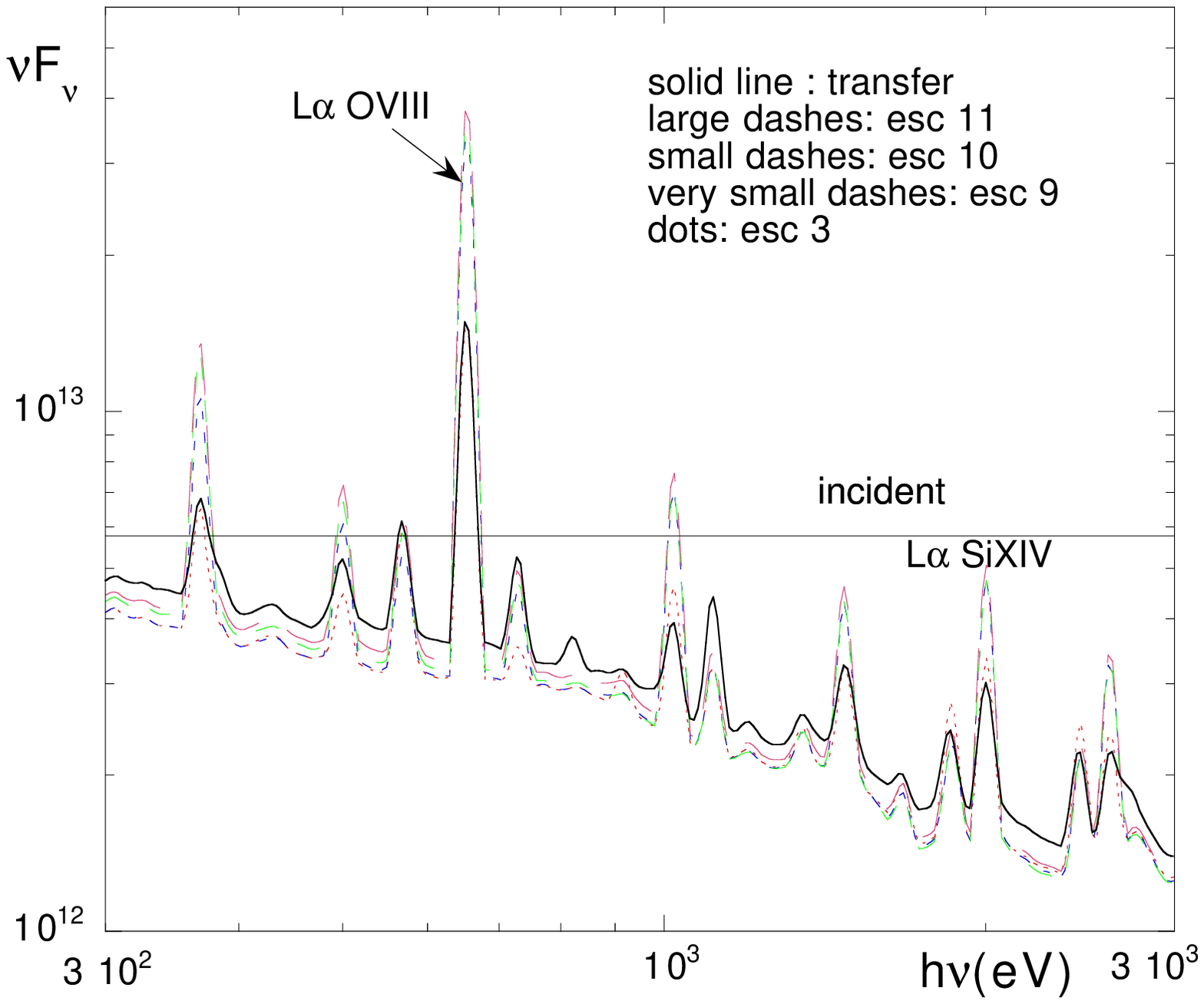,width=9cm}
\psfig{figure=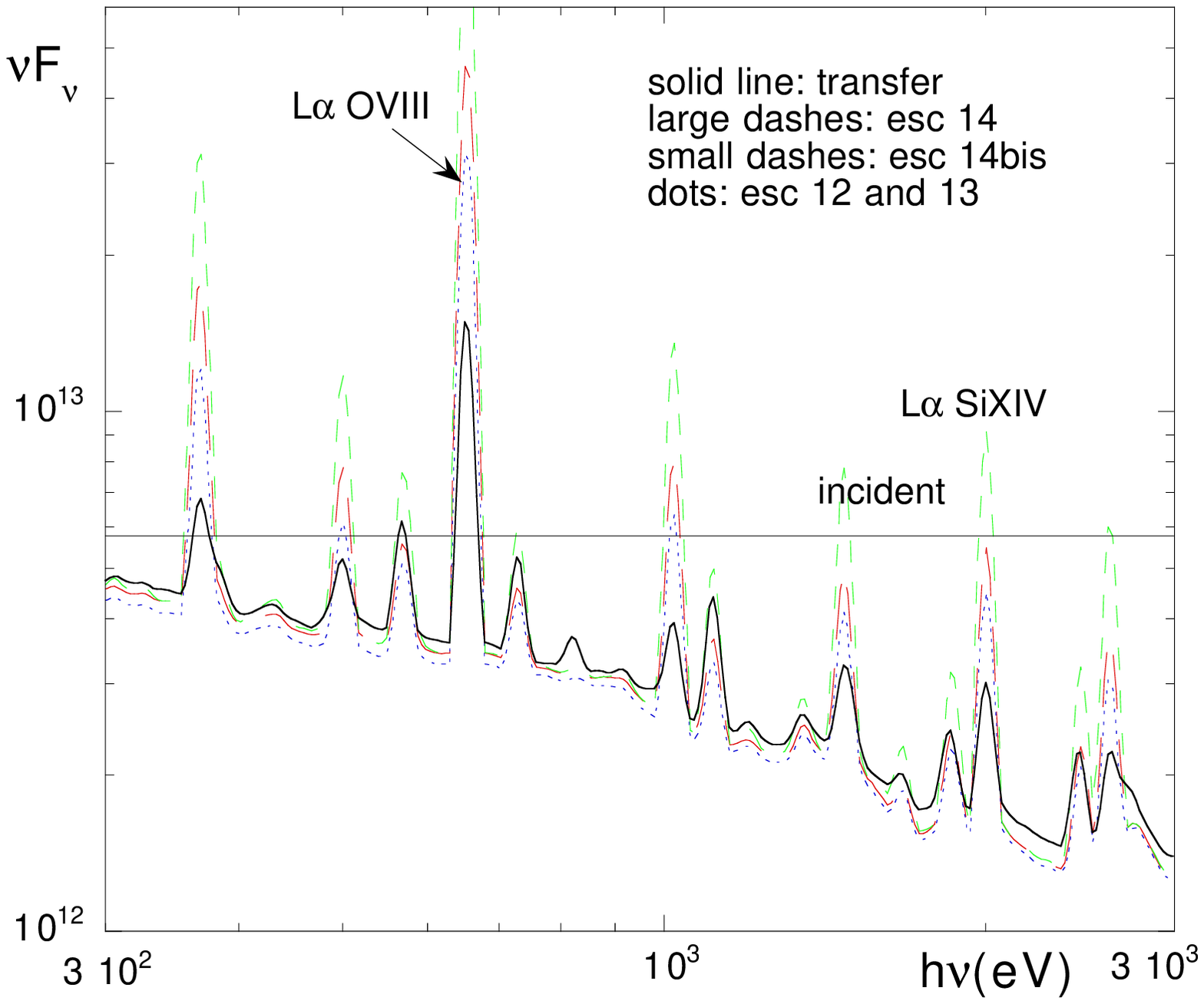,width=9cm}
\caption{Reflected spectrum in ergs cm$^{-2}$ s$^{-1}$,  for the reference 
model,
with the different approximations and 
with the full transfer computation. 
 The spectral resolution is 30. }
\label{fig-spe}
\end{center}
\end{figure}

\begin{figure}
\begin{center}
\psfig{figure=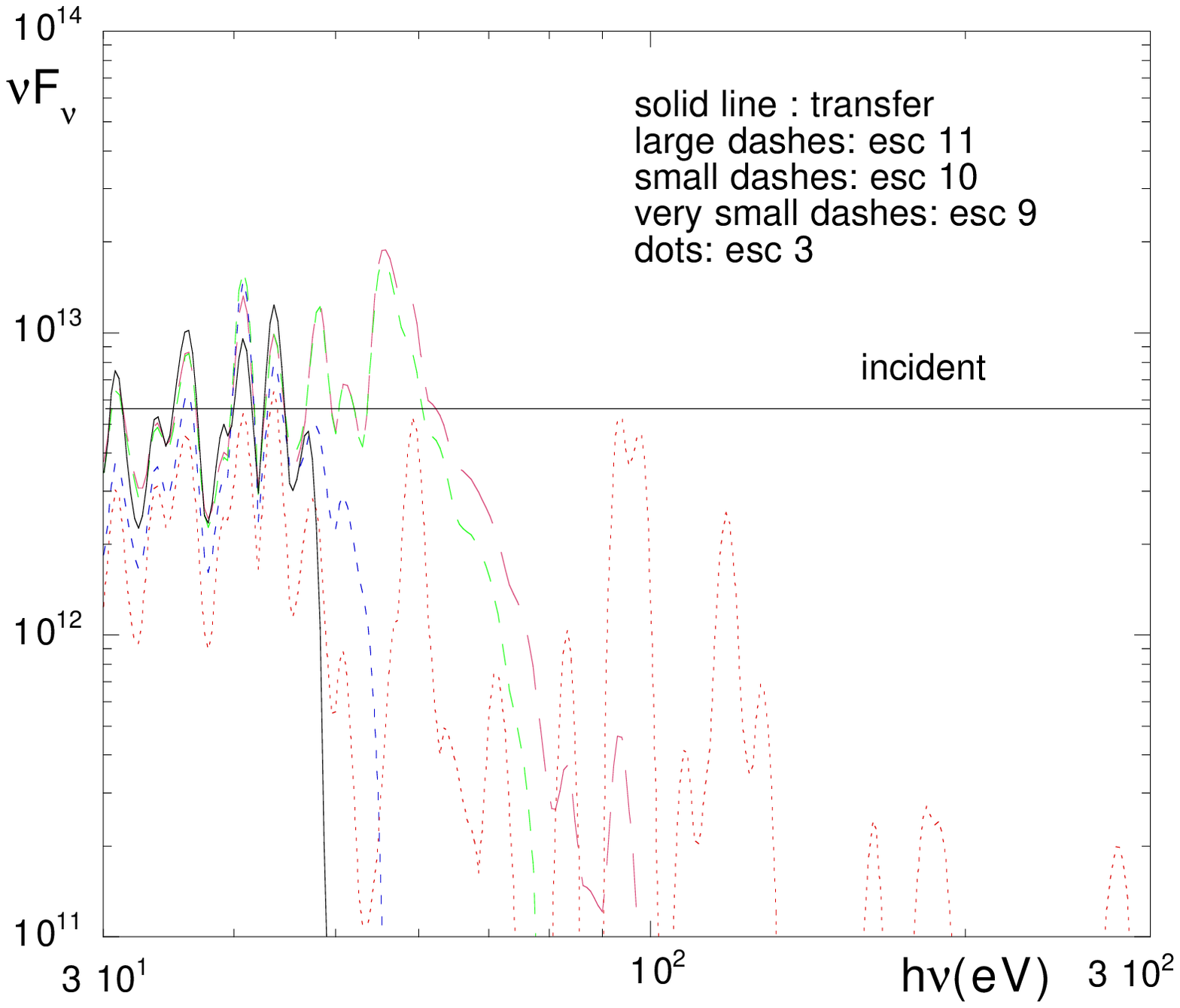,width=9cm}
\psfig{figure=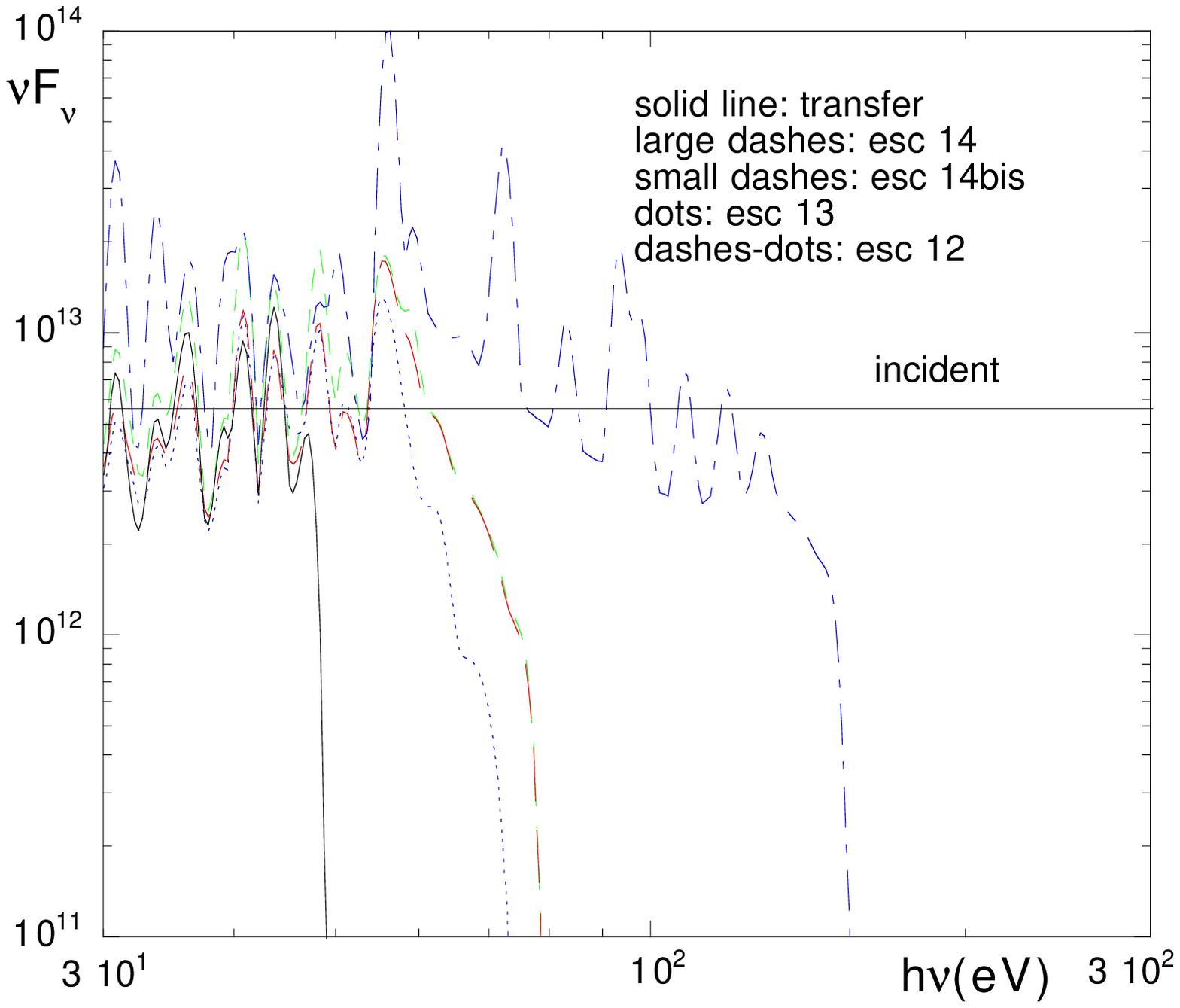,width=9cm}
\caption{Outward emitted spectrum in ergs cm$^{-2}$ s$^{-1}$, 
for the 
reference model,
with the different approximations and 
 with the full transfer computation. 
 The spectral resolution is 30. The flux is always overestimated with the 
 approximations, owing to the overestimation of line ionization and heating in the 
 deepest layers.}
\label{fig-spebis}
\end{center}
\end{figure}

\begin{figure}
\begin{center}
\psfig{figure=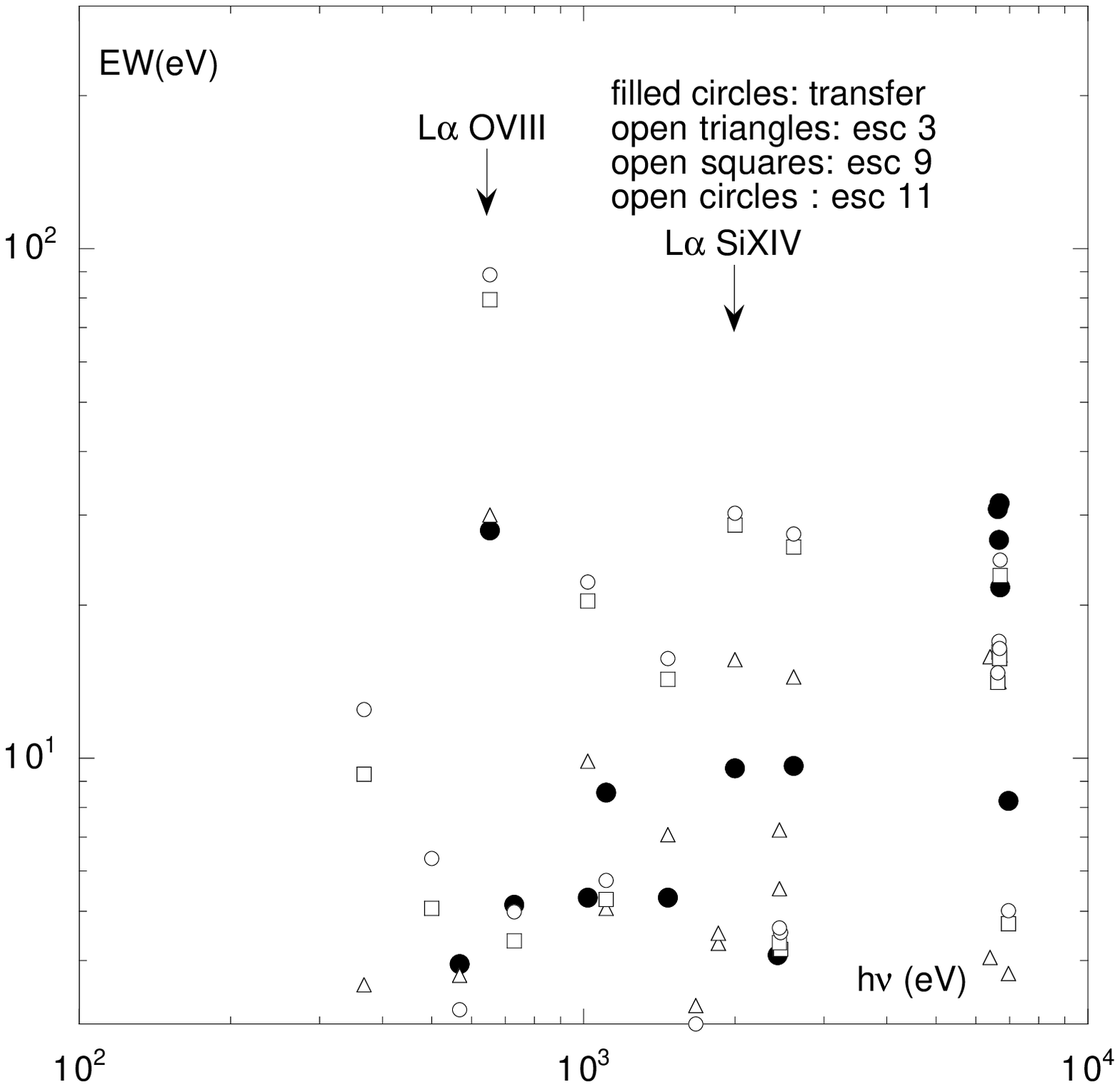,width=9cm}
\psfig{figure=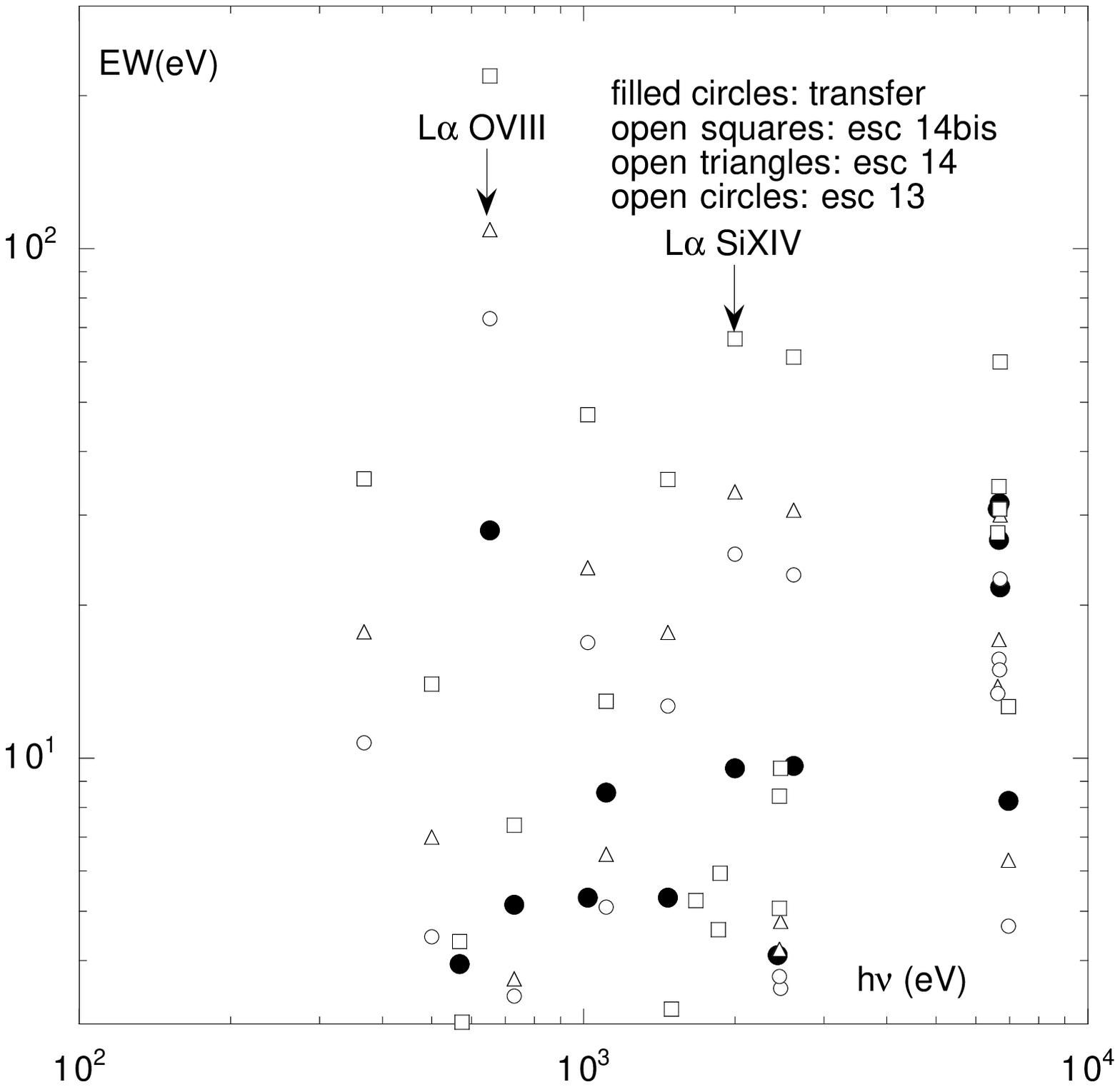,width=9cm}
\caption{Line equivalent widths in eV, with respect to the total
continuum (reflected plus incident), for the 
reference model,
with the different approximations and 
 with the full transfer computation. Some 
 approximations lead to an overestimation of the EWs by one order of 
 magnitude. Escape 3, which seems to be the best, gives weak lines 
 because it overestimates continuum emission through the back side.The 
 other approximations lead to at least a factor of three overestimation 
 of the intense lines. }
\label{fig-EWraies}
\end{center}
\end{figure}

\begin{figure}
\begin{center}
\psfig{figure=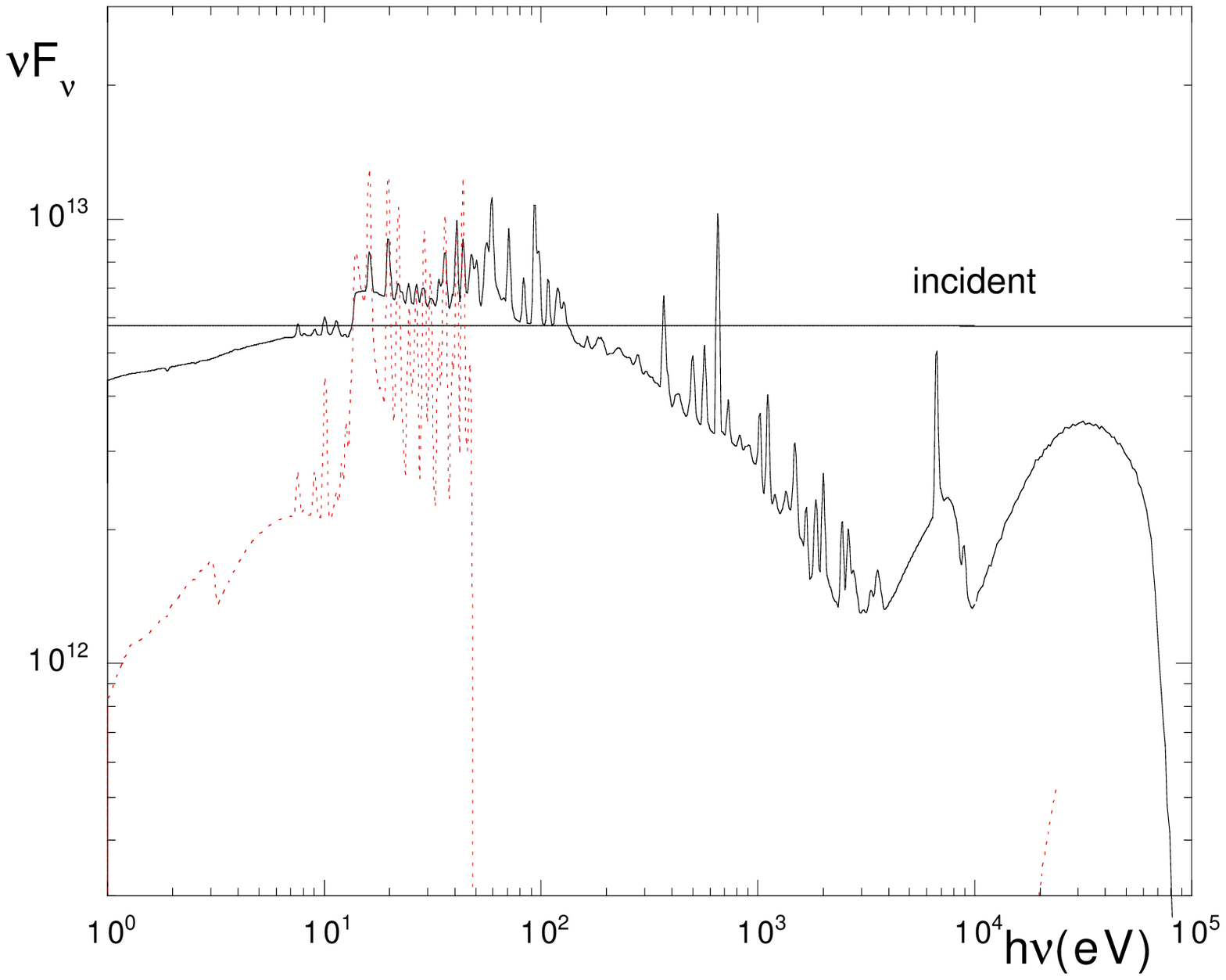,width=9cm}
\psfig{figure=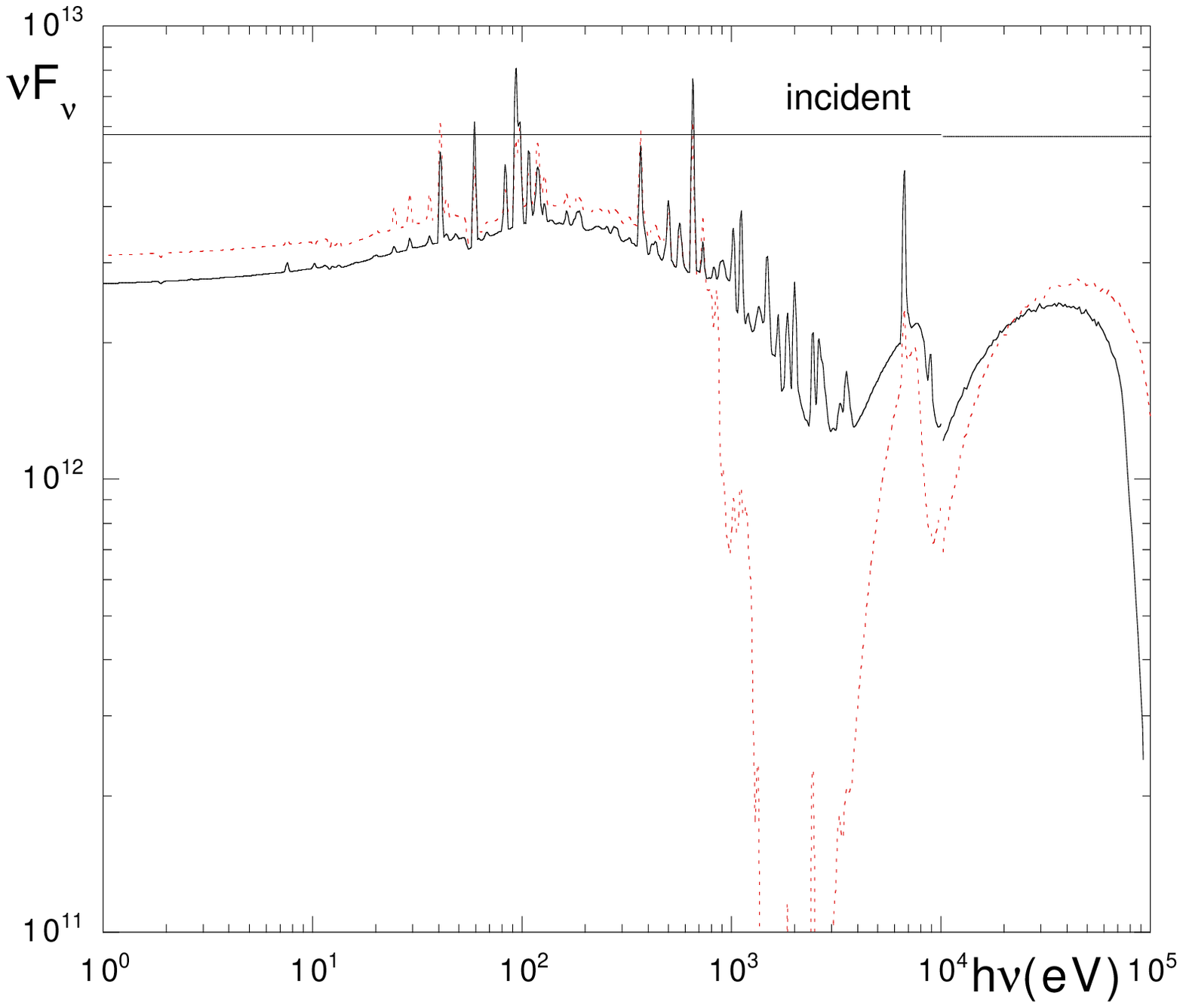,width=9cm}
\caption{The reflected (solid line) and outward emitted (dotted line) 
spectra in ergs cm$^{-2}$ s$^{-1}$, 
 with the full transfer computation:
  top panel: the reference model; bottom 
 panel: the thin model. The spectral resolution is 30.}
\label{fig-speter}
\end{center}
\end{figure}

 \begin{figure}
\begin{center}
\psfig{figure=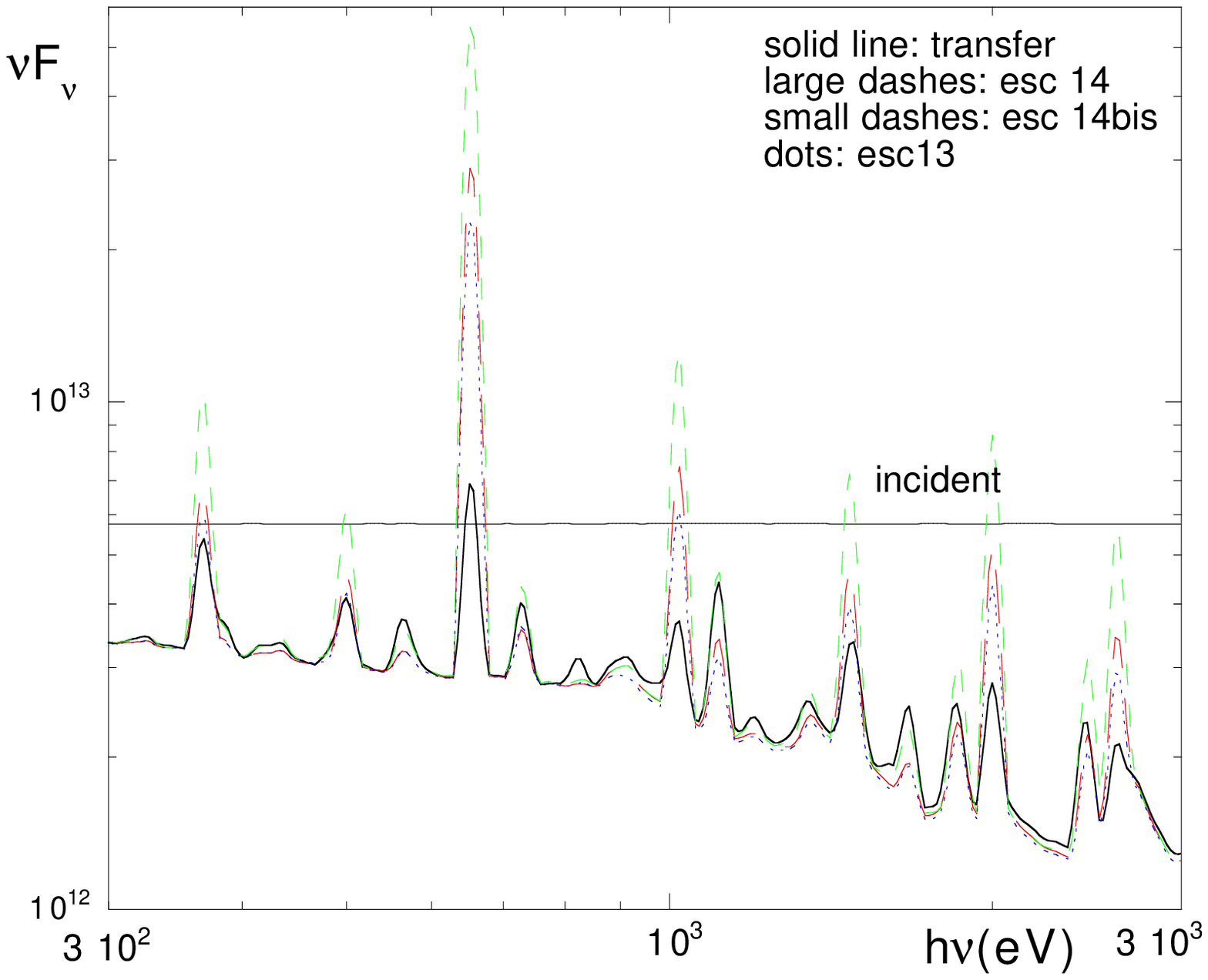,width=9cm}
\psfig{figure=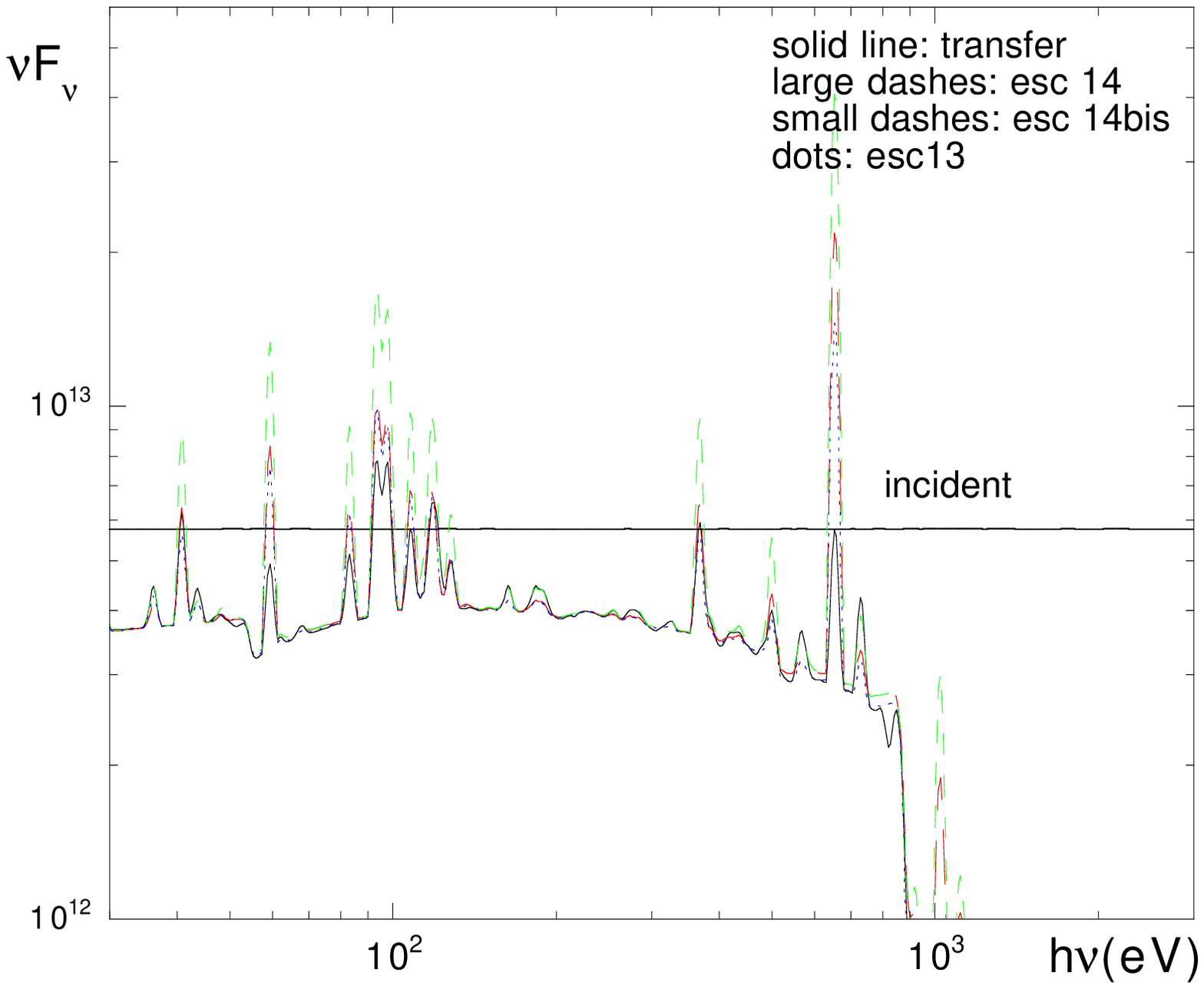,width=9cm}
\caption{Reflected (top panel) and outward emitted (bottom panel) spectrum 
in ergs cm$^{-2}$ s$^{-1}$, for the thin model,
with different approximations and 
 with the full transfer computation. The spectral resolution 
 is 30. The outward emitted spectrum is much better computed than in 
 the case of the thicker slab, but still the lines are strongly 
 overestimated.}
\label{fig-c24-spe}
\end{center}
\end{figure}

\subsection{Results: the spectrum}

Figs. \ref{fig-spe} and \ref{fig-spebis} 
display respectively the reflected and the 
outward emitted spectrum, in the full transfer 
and in the approximate computations. They are not intended to allow a 
detailed 
comparison, but only to show the broad differences between the 
results.

 We first see that the reflected continua (Fig. \ref{fig-spe}) are very similar in all 
computations. This was expected as the structure in temperature and 
ionization of the hot part of 
the irradiated atmosphere is similar.

The outward emitted continua (Fig. \ref{fig-spebis}) differ considerably
 in the UV and EUV ranges: it is due to the different
ionization structure of the ``cold" layers. With the transfer treatment the spectrum 
 is completely cut above 54 eV, 
owing to the presence of a large fraction of He$^{+}$ ions when
$\tau_{es}\ge 3$, while with the escape probability 
approximations, helium remains completely ionized up to the back of the 
slab. It is particularly obvious 
for Escape 3
 approximation, because the lines emitted by the surface layers are 
not reabsorbed by the continuum in the cold layers (note that Escape 3 is made for 
a semi-infinite medium). Escape 12  gives also a completely wrong 
outward spectrum, corresponding to the large mismatch in the energy 
balance.

  Fig. \ref{fig-EWraies} displays the equivalent widths in eV of the most 
  intense lines, with respect to the total
continuum (reflected plus incident - the outward emitted continuum is negligible 
in this frequency range), for the different 
 approximations and for the transfer treatment. Strong discrepancies 
appear
 between the different escape approximations, and we see that 
 basically 
 they all lead to overestimates of the equivalent widths, sometimes by more than 
 one order of magnitude.
 Note for instance the large value of the EW of 
the OVIII 
L$\alpha$ line at 650 eV with Escape 14 (220 eV), while it is only 
28 eV with the transfer 
treatment. 
 Note also that in the blend of iron lines near 
7 keV, some lines are underestimated by the approximations, and other 
are overestimated, but they are not very different from the transfer 
treatment.  Curiously, Escape 14bis, which is in 
 principle the best approximation, give the worst results: the 
 lines are in average 5-10 times more intense than with the transfer 
 treatment, while this factor is only about 3 for the other 
 approximations.
 Escapes 11 and 12 give exactly the same line spectrum, meaning that 
 the approximation chosen for $P_{line}$ is actually not the dominant 
 factor for the reflected spectrum, and that it is the treatment of the 
 continuum absorption and of 
 the Thomson scattering which plays the crucial role. The best 
 approximation for the reflected lines is Escape 3, which is the worse for the 
 outward spectrum! It is because the energy balance for 
 this approximation is strongly negative, which is itself due to the 
 fact that a 
 large fraction of the radiation escapes from the back of the slab. On 
 the other hand this approximation is the only one that takes into 
 account Thomson-Compton scattering in the statistical equations, and we will 
 see that it is in this respect better than the others.
 
 We have not shown here the hard X-ray spectra:  they 
 are comparable with the escape approximations and with the transfer 
 treatment. It is due to the similarity of the
 thermal state of the surface layers (cf.
 Fig. \ref{fig-T}). For illustration, 
 Fig. \ref{fig-speter} (top panel) displays the whole reflected and outward 
 emitted spectrum for the 
 full transfer computation.
 
 \subsection{A ``thin" model}

 To check whether the results would be better for the escape 
 approximations with a thinner slab, we have run a model with a column 
 density equal to $10^{24}$ cm$^{-2}$ (i.e. $\tau_{es}=0.8$). All the other 
 parameters are identical to those of the reference model.
 
 In this case the 
 medium is almost homogeneous, with a temperature of the 
 order of 
 1.5 10$^{6}$ K in the whole slab. The escape probability 
 approximations should therefore be more valid. It is only partly the 
 case. The corresponding spectra 
 are displayed on Fig. \ref{fig-c24-spe} for Escape 13, 14, and 14bis 
 approximations. The agreement with the full 
 transfer computation is indeed much better for the outward emitted 
 continuum, but it is still very bad for the lines. Note in particular again the 
 strong discrepancy of the OVIII L$\alpha$ line. It is due to the fact 
 that in both the ``thin" and the ``thick" cases, the X-ray 
 lines are formed in the same region $\tau_{es} \le 1$, and the 
 transfer suffers from the same problems, which will be 
 discussed in the next section.

Fig. \ref{fig-speter} (bottom panel) displays the whole reflected and outward 
 emitted spectra for the 
 full transfer computation, for this model.
 
 \section{Why are the line intensities larger with the escape approximations 
than with the full transfer treatment?}

 \begin{figure}
\begin{center}
\psfig{figure=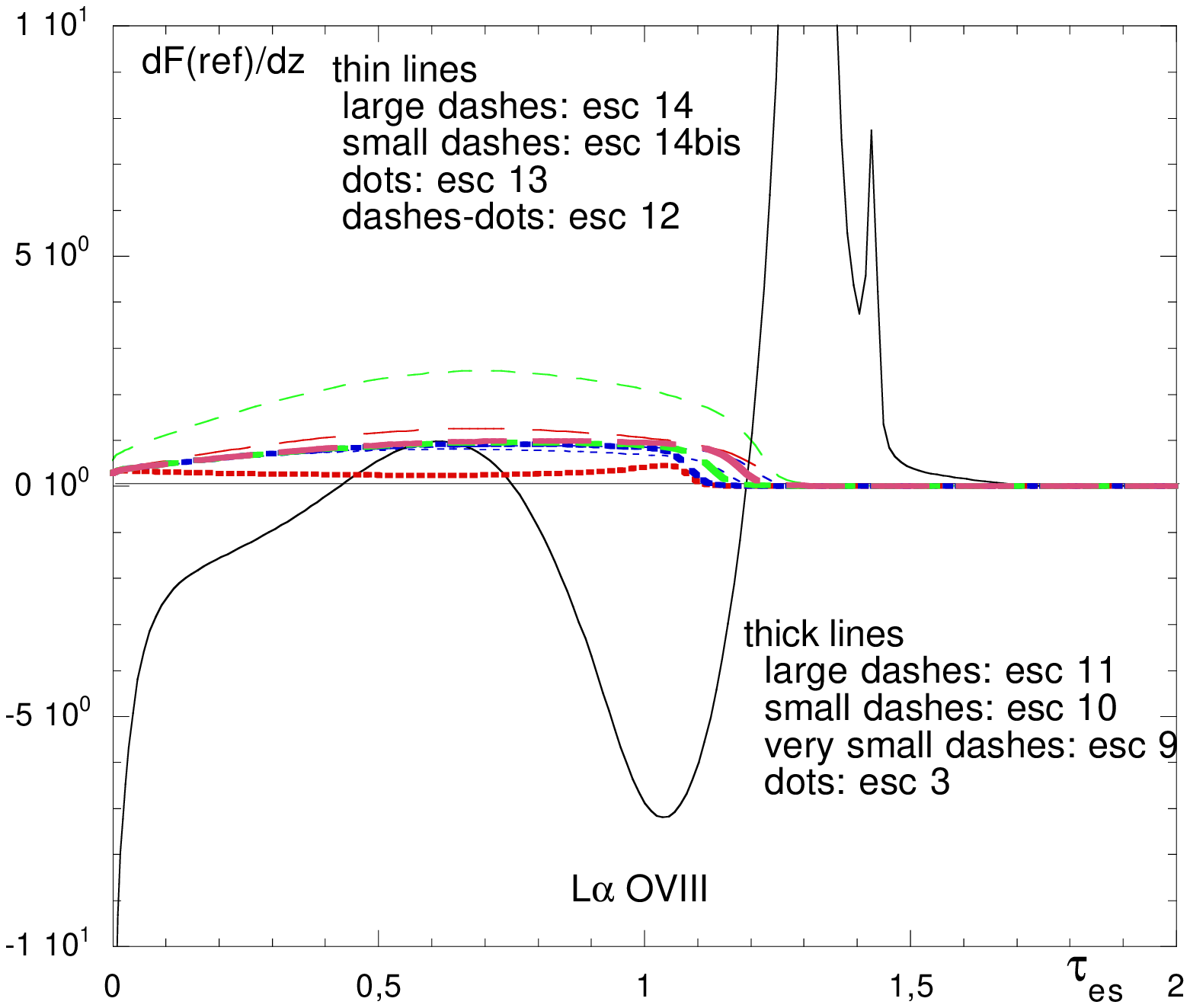,width=9cm}
\psfig{figure=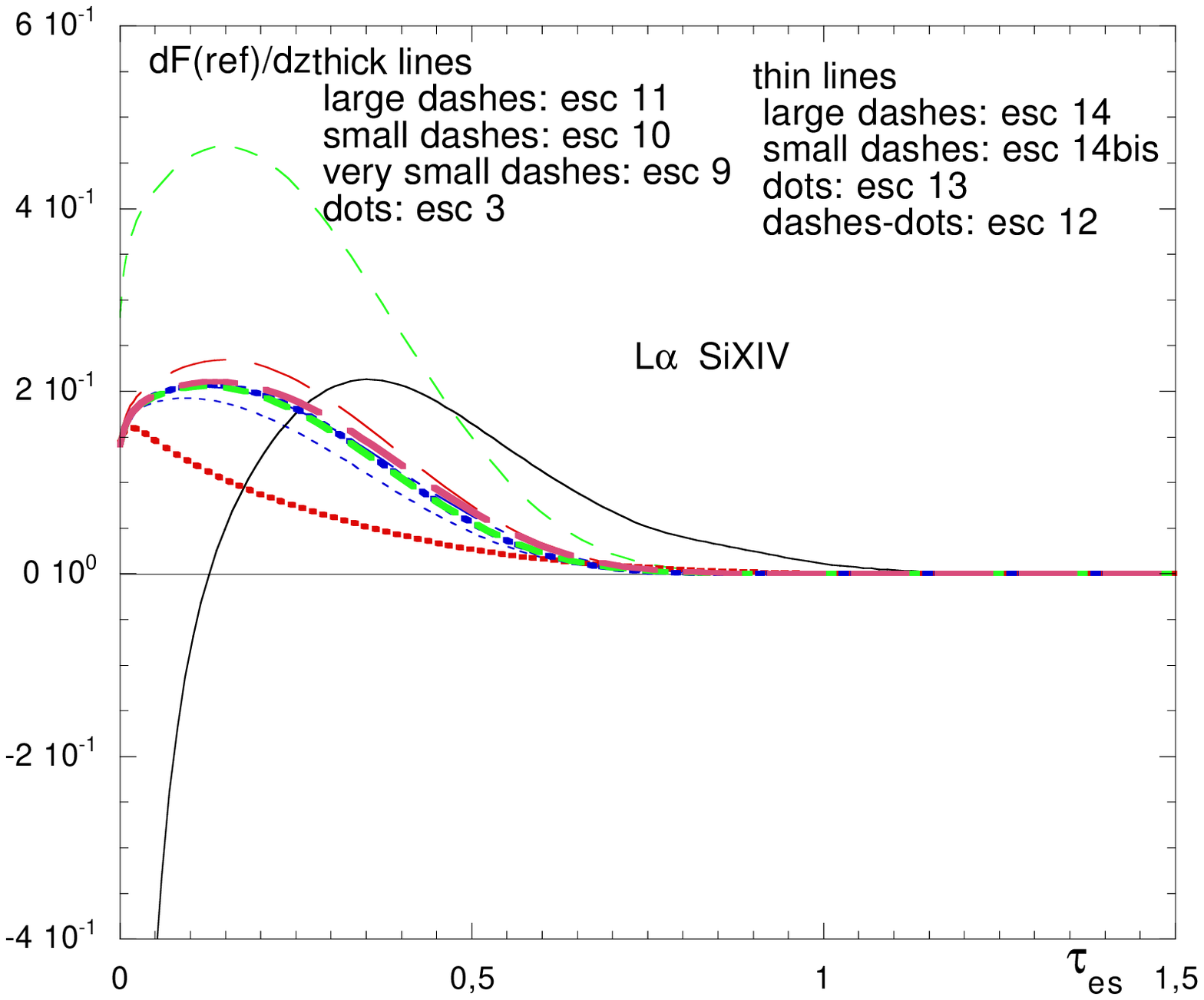,width=9cm}
\caption{The differential reflected flux versus depth, for OVIII L$\alpha$
 and SiXIV L$\alpha$ as a function of  $\tau_{es}$, for the reference model
  with the different 
 escape approximations. The solid line gives the value obtained with the full 
 transfer treatment. We notice that the results have almost nothing in 
 common, the transfer treatment leading to large negative values, 
 which do not exist with the escape approximations, and which explain 
 the overestimation of the line fluxes. The smaller line fluxes 
 obtained with Escape 3 are actually due to the local 
treatment.}
\label{fig-Fz}
\end{center}
\end{figure}

  \begin{figure}
\begin{center}
\psfig{figure=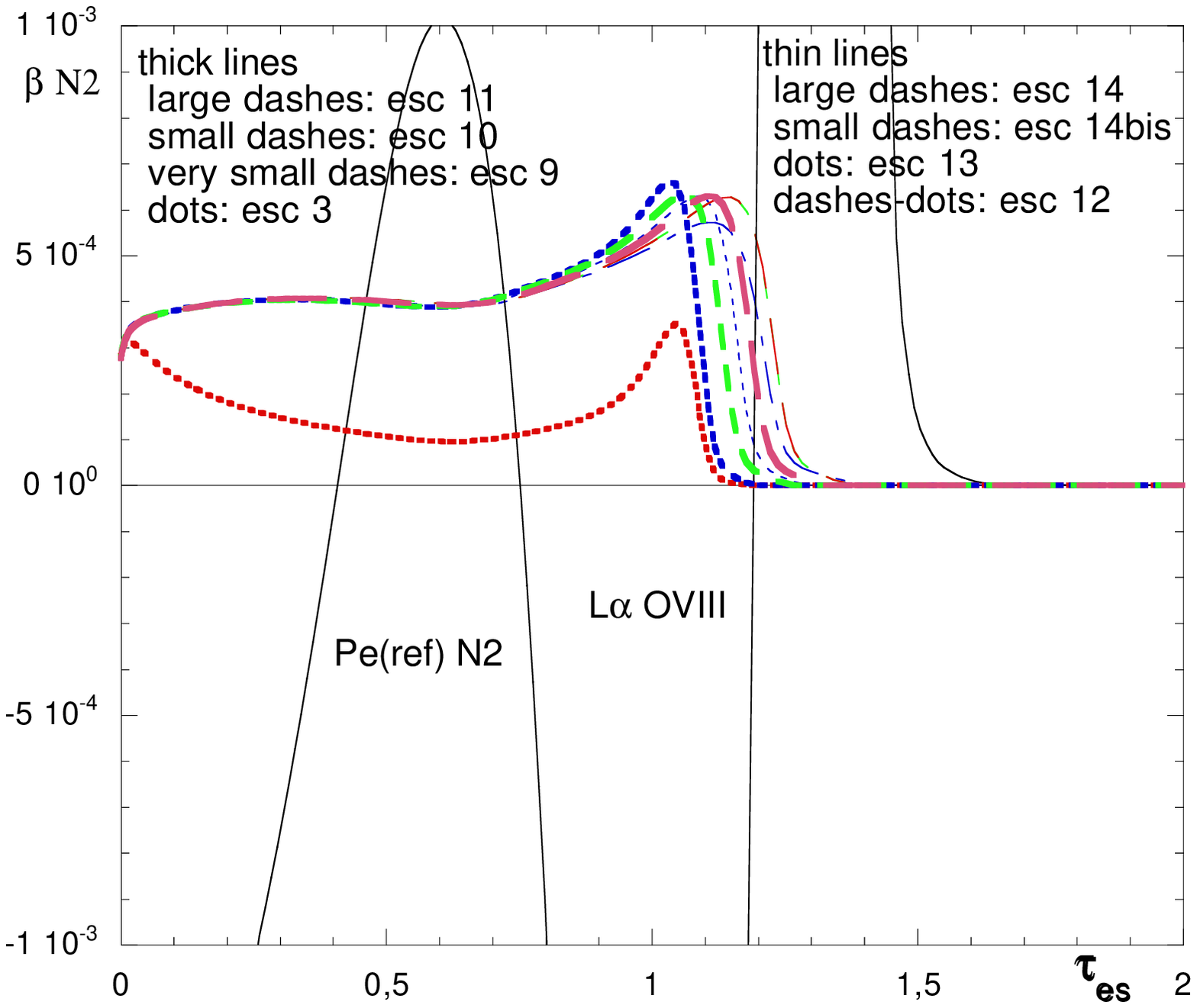,width=9cm}
\psfig{figure=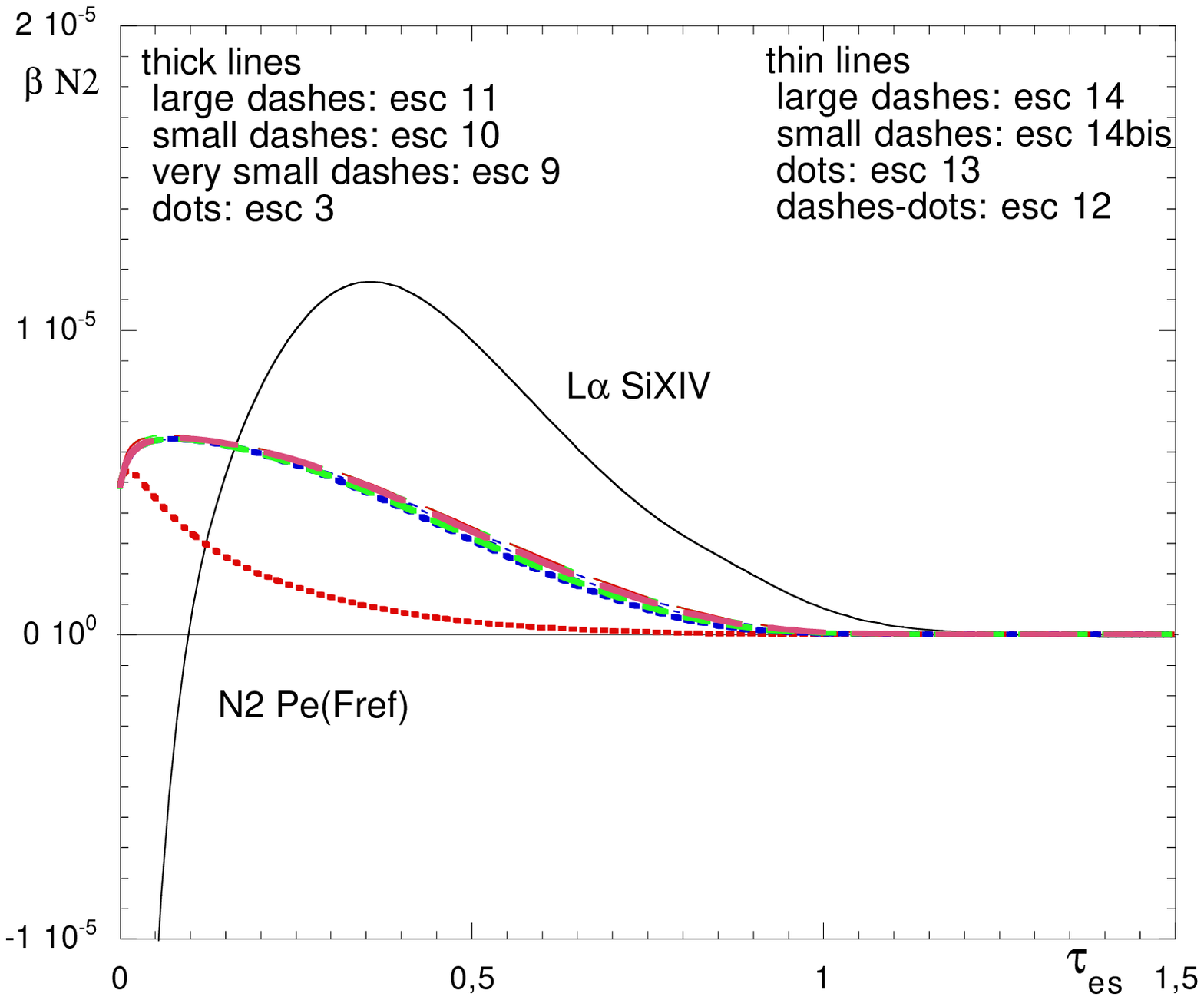,width=9cm}
\caption{The product $\beta(ref)\times n_{2}$, proportional to the 
local line 
emissivity, as a function of $\tau_{es}$, for the reference model
  with the different 
 escape approximations. The solid line gives the product $P_{e}(ref)\times 
 n_{2}$, with $P_{e}(ref)$ defined in the text for the transfer 
 treatment. Again the results are very different, with strong 
 negative values in the case of the transfer treatment. The smaller 
 value of the product with Escape 3 explains why the reflected fluxes are 
 smaller. }
\label{fig-n2beta}
\end{center}
\end{figure}

  \begin{figure}
\begin{center}
\psfig{figure=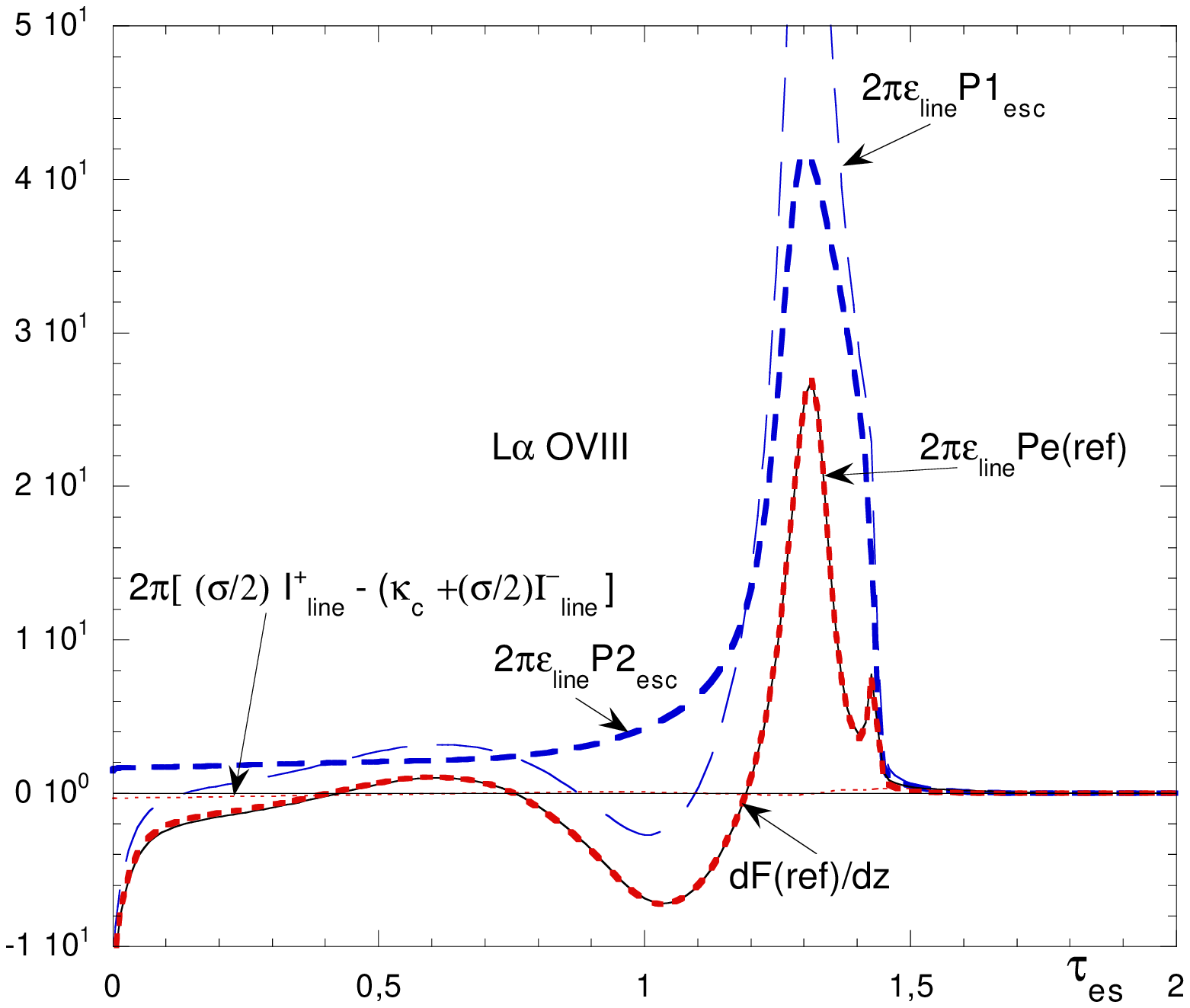,width=9cm}
\psfig{figure=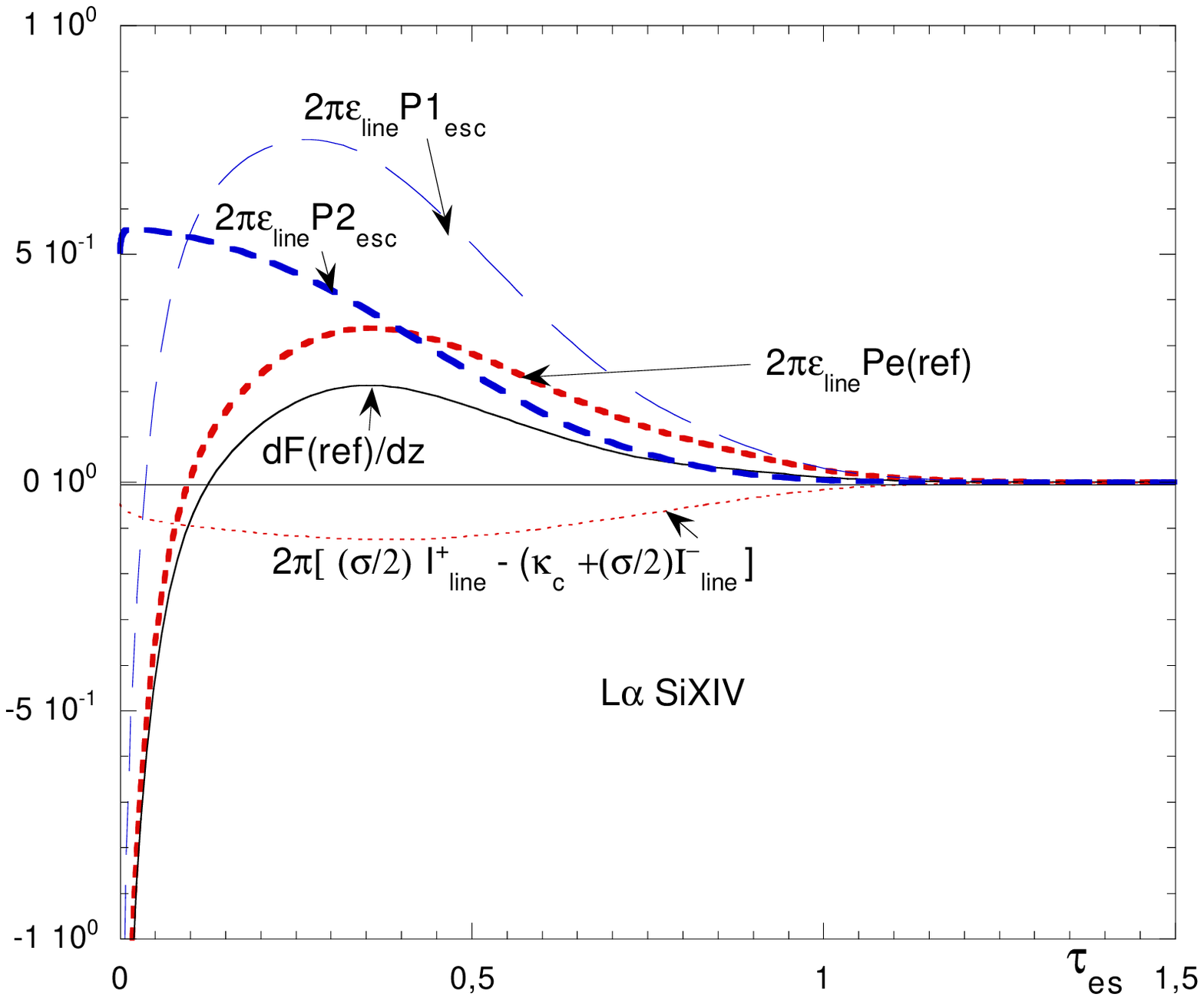,width=9cm}
\caption{The different parts of the right member of Eq. 41,
 as functions of $\tau_{es}$, for the reference model, 
with the transfer treatment and for L$\alpha$ OVIII and SiXIV. 
It shows that for both lines (it is even 
more obvious for L$\alpha$ OVIII $), {dF(ref)/ 
dz}$ is  determined by $P_{e}(ref)$. $P_{e}(ref)$ itself 
is the difference of the two terms $P1_{esc}$ and $P2_{esc}$. The 
figure
shows that the negative values 
of $P_{e}(ref)$ (and of $dF_(ref)/dz$) are due to the presence 
of the term $P2_{esc}$, i.e. to excitations by the 
diffuse continuum.  }
\label{fig-dFrsdz-decomp}
\end{center}
\end{figure}

We would like to know whether the line intensities differ as a 
consequence of the local treatments, or of the treatments of 
the photons during their travel towards the surface, or both. It is 
not easy, because in the case of the transfer, 
local and non-local
photons are not distinguishable.

To study the problem, we have chosen to
examine two intense X-ray lines formed in different layers, 
 OVIII L$\alpha$ and at 650 eV and SiXIV L$\alpha$ at 
2 keV.

First it is necessary to know where these lines are formed.
Note that the emission of both lines is restricted to the reflected 
flux. The differential 
reflected
flux $dF_{ref}/dz$ is given by Eq. \ref{eq-Fref} for Escape 14, and 
one has
equivalent expressions for the other escape approximations. For the 
transfer treatment, one cannot use directly Eqs. 
\ref{eq-flux-line-integre} or
\ref{eq-flux-line-integrebis}, because they give the flux in both 
directions, while we want only the reflected flux. In the two-stream 
approximation,
the transfer equation becomes:

\begin{eqnarray} {1 \over \sqrt{3}}{dI_{\nu}^{+} \over dz} &=& -(\kappa_{\nu} 
+{\sigma \over 2})I_{\nu}^{+}+{\sigma \over 2}I_{\nu}^{-}+\epsilon_{\nu} \\
\nonumber {-1 \over \sqrt{3}}{dI_{\nu}^{-} \over dz} &=& -(\kappa_{\nu} +{\sigma 
\over 2})I_{\nu}^{-}+{\sigma \over 2}I_{\nu}^{+}+\epsilon_{\nu} 
\label{eq-trans-2stream}
\end{eqnarray}

where $I_{\nu}^{-}$ and $I_{\nu}^{+}$ are the intensities towards the 
surface  and towards the 
back, and $F_{\nu}= {2\pi/\sqrt{3}}(I_{\nu}^{+}-I_{\nu}^{-})$. 
The reflected flux integrated on the line profile can be obtained from 
these equations, like Eq. \ref{eq-flux-line-integre} was 
obtained from Eq. \ref{eq-trans-angle}:

\begin{eqnarray}
&& {dF_{line}(ref)\over dz}={2\pi\over \sqrt{3}}{dI^-_{line}\over dz}=
\\ \nonumber
&&  \left[ 2\pi 
\epsilon_{line}-2\pi\kappa_{line} \int \phi_{\nu}I^-_{\nu}d\nu  \right] 
\\ \nonumber
&& -2\pi(\kappa_{c}+{\sigma\over 2})I^-_{line} + 2\pi {\sigma\over 
2}I^+_{line},
\label{eq-dFrefsdz}
\end{eqnarray}
where $I^-_{line}$ and $I^+_{line}$ are integrated on the line
($I_{line}=\int  (I_{\nu}-I_{c})d\nu$), and $\int 
\phi_{\nu}I^-_{\nu}d\nu$ takes into account the continuum intensity 
$I^-_{c}$.

Fig. \ref{fig-Fz} shows $ {dF_{line}(ref)/ dz}$ for the two lines, as 
a function of the optical depth, for a few escape approximations and 
for the full transfer treatment. Actually there is almost nothing in common 
between the results of the two treatments! It is obvious that the smaller values 
of the line intensities with the transfer treatment come from the fact that 
 $ {dF_{line}(ref)/ dz}$ is negative
in a large fraction of the medium, while it is not the case 
in the escape treatment. This is due to the fact that the flux is 
directed towards the surface or towards the back, according to the 
variation of the source function (see below).  Finally, one should note 
the much smaller flux obtained with Escape 3.

Note also that with the transfer treatment, the emission of OVIII L$\alpha$
 is provided 
by a small layer located at  $\tau_{es}\sim 1.3$.
We have seen on Fig. \ref{fig-ion-O} that it is precisely the region 
where OVIII ions are dominant.  Thus we can 
infer that  the  emission is mainly due to excitation and not to 
recombination. Actually it is excitation by the diffuse continuum 
flux, and not by collisions.
SiXIV  emission is important  up to $\tau_{es}\sim 1$, again when the 
SiXIV
ions dominate, 
and SiXIV  L$\alpha$ is also dominated by radiative excitation. 

One can go a step further in our understanding. 
In analogy with Eq. \ref{eq-flux-line-integrebis}, 
one can write 
Eq. 40:

\begin{eqnarray}
&& {dF_{line}(ref)\over dz}=  2\pi 
\epsilon_{line} \times P_{e}(ref) +
\\ \nonumber
&& -2\pi(\kappa_{c}+{\sigma\over 2})I^-_{line} + 2\pi {\sigma\over 
2}I^+_{line},
\label{eq-dFrefsdz}
\end{eqnarray}
where
\begin{equation}
P_{e}(ref)={-2\pi\kappa_{line} \int \phi_{\nu}I^-_{\nu}d\nu + 2\pi 
\epsilon_{line}\over 2\pi  
\epsilon_{line}}
\label{eq-trans200}
\end{equation}
is the equivalent of an escape probability for the reflected flux.

Fig. \ref{fig-n2beta} displays the product  $P_{e}(ref)\times 
 n_{2}$, and compares it to  $\beta(ref)\times n_{2}$ 
for different escape approximations ($\beta$ is given by Eq. 
\ref{eq-beta}). This last product is proportional to the 
local line 
emissivity with the escape approximations. Again there is little in 
common between the two treatments, but now we see that the reflected fluxes 
are determined by the local treatment, and not by the non-local one. 
This is in particular the case of Escape 3, which is the only local 
treatment to take into account Thomson-Compton scattering 
in the statistical equations (see below).

The second part of the right member of Eq. 41
 is 
negligible, or at least not dominant, so the behavior of ${dF_{line}(ref)/ 
dz}$ is  determined by that of $P_{e}(ref)$. $P_{e}(ref)$ itself 
is the difference of two terms, 

\begin{eqnarray}
P1_{esc}&=&{-2\pi\kappa_{line} [\int \phi_{\nu}I^-_{\nu}(line)\ d\nu] + 2\pi 
\epsilon_{line}\over  2\pi 
\epsilon_{line}} \\ \nonumber
&=& 1\ -\ {\int \phi_{\nu}I^-_{\nu}(line)\ d\nu\over S_{line}}
\label{eq-trans300}
\end{eqnarray}
which can be identified with the one side escape probability 
computed with Eqs. \ref{eq-escape00} and \ref{eq-esc-1}, and a second term:  

\begin{equation}
P2_{esc}={2\pi\kappa_{line} I^-_{c} \over 2\pi 
\epsilon_{line}}\ = \ {I^-_{c}\over S_{line}}
\label{eq-trans400}
\end{equation}
which corresponds to {\it excitation of the line by the diffuse 
continuum, which is not taken 
into account in the expression of the escape probability}. 

Fig.\ref {fig-dFrsdz-decomp} shows $2\pi \epsilon_{line}P1_{esc}$ and 
$2\pi \epsilon_{line}P2_{esc}$. It proves that {\it the negative values 
of $P_{e}(ref)$ and $dF_{line}(ref)/dz$ are mainly due to the 
diffuse continuum}. It is a confirmation that the problem is mainly a local 
one, and does not depend much on the way the photons are treated 
non-locally in the escape approximations\footnote{Note that these 
computations should be performed in double precision, 
to compute correctly differences between nearly equal quantities}.

  \begin{figure}
\begin{center}
\psfig{figure=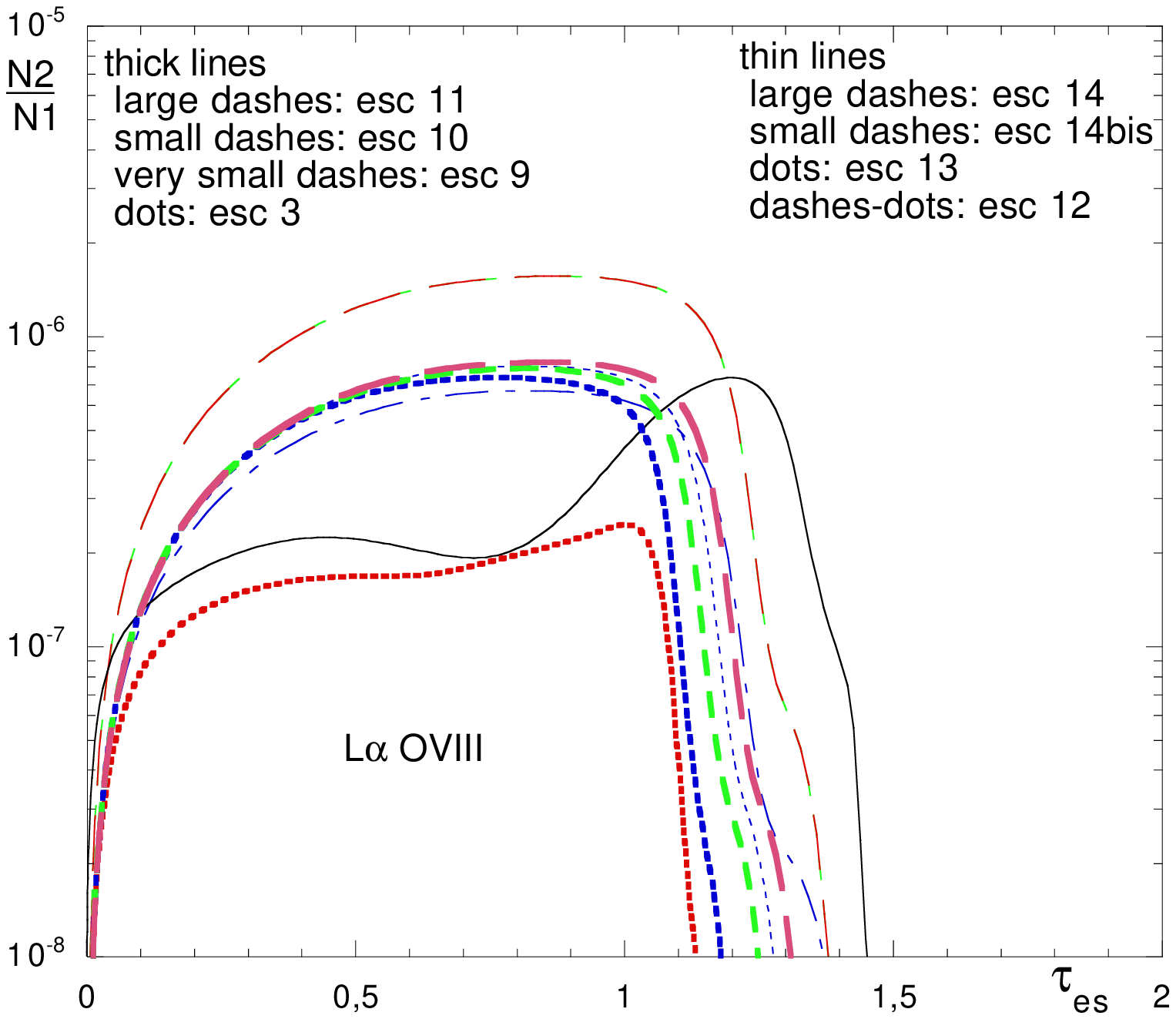,width=9cm}
\psfig{figure=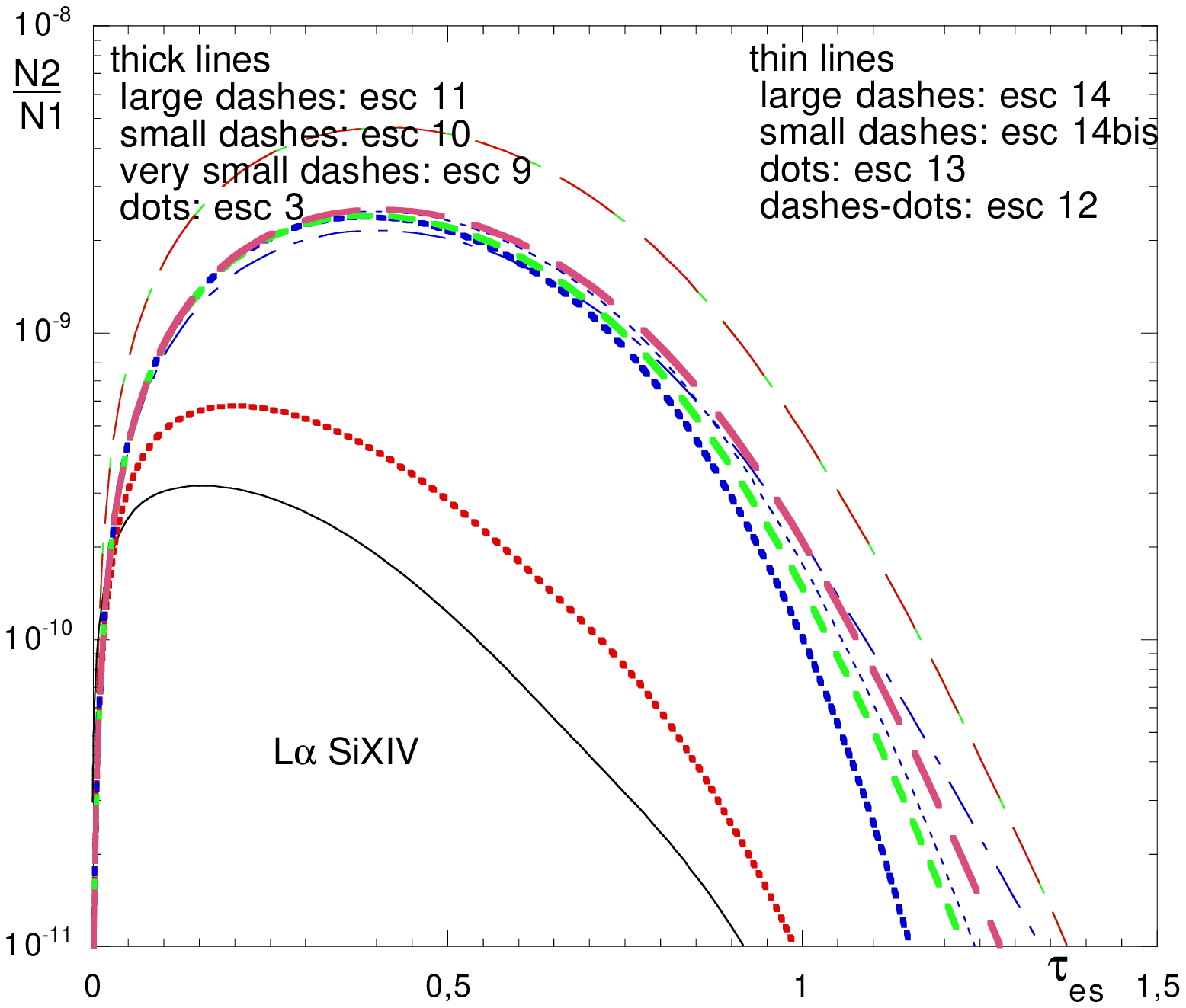,width=9cm}
\caption{The ratio 
of the population of the upper level to that of the lower level, 
$n_{2}/n_{1}$ (proportional to the source function), for OVIII L$\alpha$
 and SiXIV L$\alpha$, as a function of  $\tau_{es}$, for the reference model
 with the different 
 escape approximations. The solid line gives the ratio for the full 
 transfer treatment. Except Escape 3, the other approximations strongly overestimate 
 the source function, but Escape 3 underestimates it for the OVIII 
 line in the region 
 where it is emitted.}
\label{fig-n2n1}
\end{center}
\end{figure}

\begin{figure}
\begin{center}
\psfig{figure=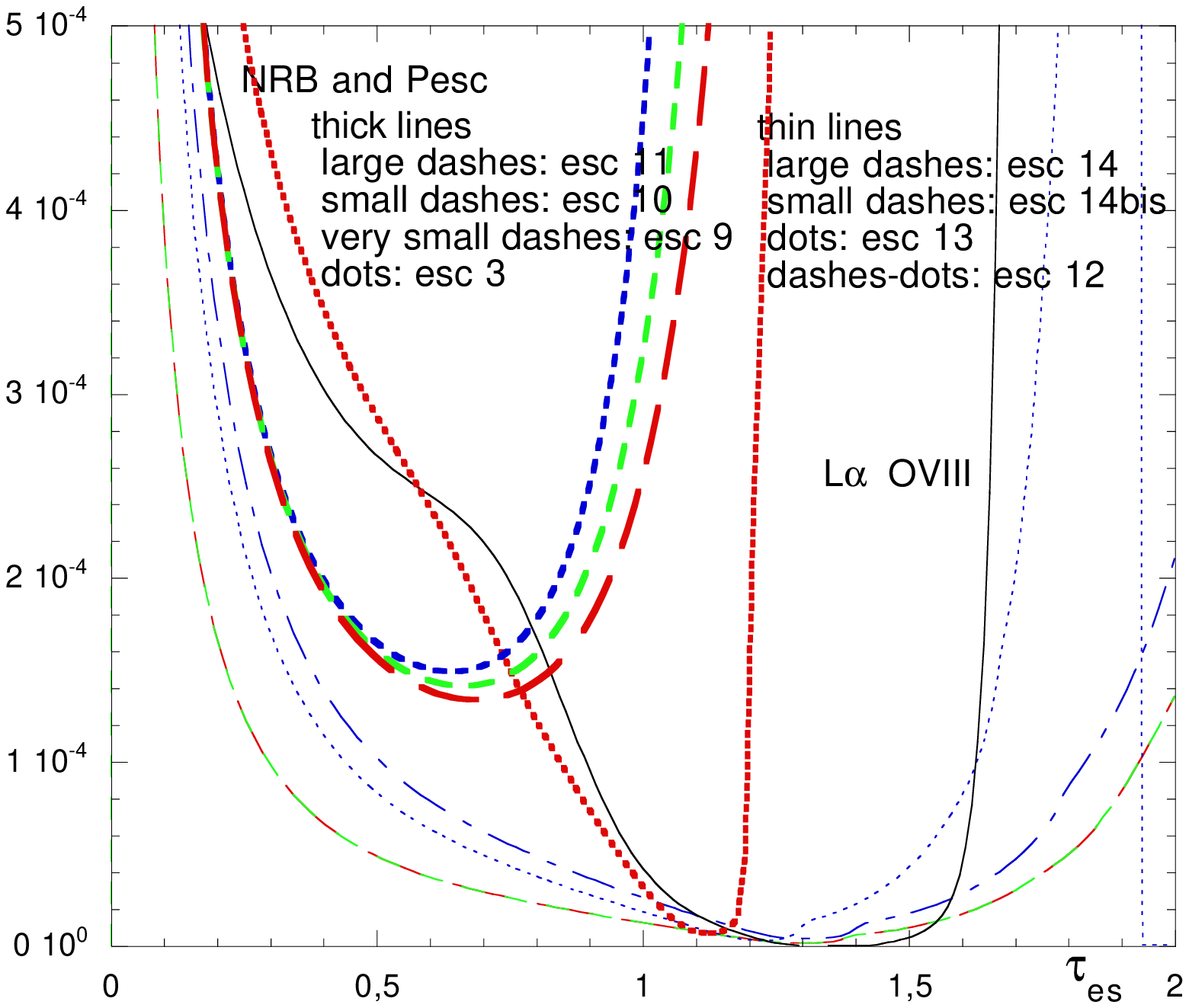,width=9cm}
\psfig{figure=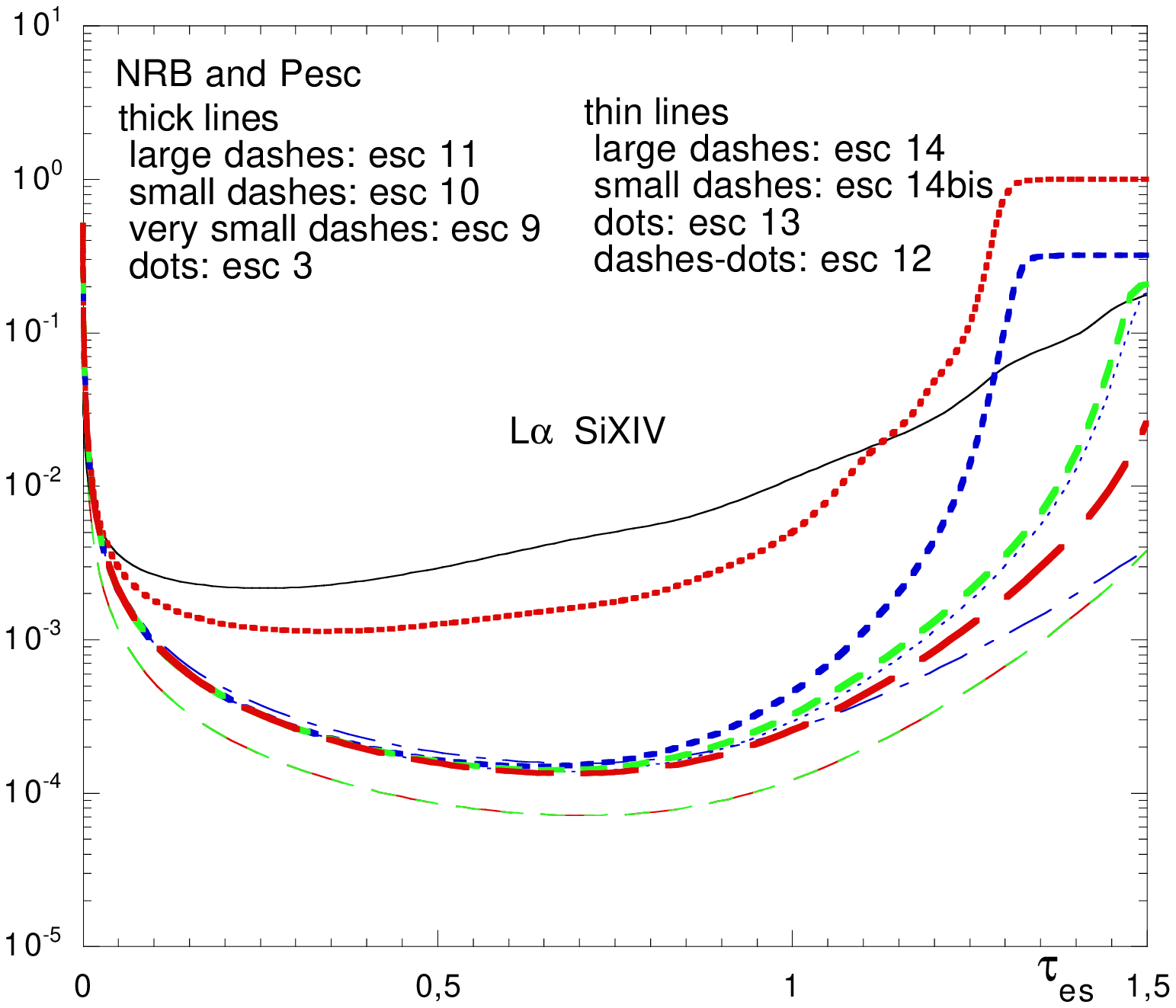,width=9cm}
\caption{Comparison between the NRB and the escape probability  $P_{esc}$ which replaces 
the NRB in the statistical equations of the levels computed with the 
escape probability, for OVIII L$\alpha$
 and SiXIV L$\alpha$, as functions of $\tau_{es}$, for the reference model
with the different 
 escape approximations. The solid line gives NRB for the full 
 transfer treatment. The escape approximations strongly underestimate 
$P_{esc}$, which means that they overestimate the importance of local diffusions.
 Escape 3 has a different behavior in this respect, because it allows 
 the photons to escape by Thomson-Compton diffusions.}
\label{fig-NRB-Pesc}
\end{center}
\end{figure}

To stress this aspect, it is interesting to compare local quantities 
obtained with both treatments, such 
as the source function. 

Fig. \ref{fig-n2n1} 
displays the ratio 
of the population of the upper level to that of the lower level, 
$n_{2}/n_{1}$, which is proportional to the source 
function, for OVIII L$\alpha$
 and SiXIV L$\alpha$,  as a function of $\tau_{es}$. For both lines the source function 
 is very badly determined by the escape approximations in the whole 
 emission region, and it differs also
 by large factors between the different approximations, up to one order 
of magnitude.  Except Escape 3, the other approximations mainly overestimate 
 the source function. Escape 3 underestimates it for the OVIII 
 line in the region 
 where it is emitted. Since $n_{2}/n_{1}$ differs by large factors 
 from the real value with the escape approximations,
  it means that any physical process  implying the 
population of the upper level  will be 
badly determined (we recall that the first levels have almost the same 
populations in all approximations, i.e. equal to the fractional abundances of 
OVIII and SiXIV). 

Finally Fig. \ref{fig-NRB-Pesc} displays the NRB and the escape probability  $P_{esc}$ 
which is used 
instead of 
the NRB in the rate equations of the levels. There are large 
differences between the transfer treatment and the escape 
approximations, which are of course linked with the differences of the 
source function. Except Escape 3, the other approximations strongly underestimate 
$P_{esc}$, which means that they overestimate the importance of local diffusions.
 Escape 3 has a different behavior in this respect, because it allows 
 the photons to escape by Thomson-Compton diffusions, but it is also 
 bad in the region where the OVIII line is produced.

However the escape probability approximations do not overestimate 
all
the lines. The lines which are most affected are those which have an 
intense underlying diffuse continuum. For the present model, it is not the case of the Fe 
lines near 7 KeV, because the diffuse continuum is relatively weak at this 
frequency. Therefore the Fe lines are computed relatively correctly by the 
approximations (within 30$\%$). It would probably be different with a very 
thick layer and a high ionization parameter. Again, we see that other models are required 
to set the limit of validity of the escape approximations. 

\section{Conclusion}

In this paper we have shown that the escape probability approximations 
lead to a strong overestimation of the line intensities, in the case 
of a highly ionized and thick medium. They also lead to an 
overestimation of the ionization state at the back of the slab, and 
finally to an overestimation of the radiation escaping from the back. 

We have shown that the 
local treatment is more important than the way absorption is taken 
into account non-locally.  We have 
 discovered that {\bf the much smaller values of the line 
 intensities obtained with the transfer treatment are due to line 
 excitations by the diffuse continuum, which are not taken into 
 account correctly with the escape approximations}.  Considering 
 the fact that the most correct expression for the emergent flux is that of Escape 14bis,
 which gives 
 the largest line intensities (up to one order of magnitude larger 
 than the full transfer treatment), it means that {\it this 
 process should be even more 
 important than suspected}. 
 From this point of view the
best approximation seems to be Escape 3, i.e. the one made by Ko and Kallman 
(1994), which includes a local 
 escape from the line core by Thomson-Compton scattering in the statistical 
 equations, but it is actually an artifact simply due to the fact 
 that it reduces the source function in allowing this escape.
 
 We stress also that the differences in the intensities of the 
 X-ray lines
between the various approximations on one hand, and between the escape 
 approximations and the transfer treatment on the other hand, are almost exclusively due 
 to the ``transfer treatment", as the structure of the hot part of the medium is 
 almost the same in all cases, owing to the fact that the transfer of 
 the continuum was perfomed exactly in the same way in all 
 computations.
 
This result leads to the conclusion that {\it not only the line 
transfer, but also the continuum transfer, should be correctly 
handled} in order to get correct line intensities. In particular, it 
is not possible to use escape probability approximations to compute the diffuse 
continuum, as it is made in some codes. 

This study was limited to one value of the ionization parameter 
 ($\xi=1000$), and to a modest Thomson thickness. 
It has shown that the escape approximation is better in the case of a 
smaller thickness only for the emission from the back side. For the 
reflected line fluxes, it is equally bad, and we are inclined to 
think that only for Thomson thicknesses much smaller than unity, the 
escape probability approximations lead to correct X-ray line 
intensities in these hot media. We suspect that for a larger thickness,
even the structure 
of the slab is badly computed. Note that the excitation 
equilibrium of the atoms is not correct. This can be important for
 denser media, where ionizations and line 
excitations can take place from excited levels. Thus it is mandatory to 
perform other comparisons in order to delineate the parameter region 
where the escape approximations can be used. 

Finally it is worth mentioning that we have performed here all the 
computations assuming complete redistribution of frequencies, but 
with ALI it is equally possible to assume partial redistribution (cf. 
Paletou \& Auer 1995), while 
this cannot be done with 
the escape formalism, unless very crude approximations are used.

\bigskip

\noindent {\bf APPENDIX A:
Treatment of Comptonization in the lines by Titan}

\medskip

The transfer equation for the lines, in the absence of Compton 
scattering, can be written in the two-stream 
approximation:

\begin{eqnarray} {1 \over \sqrt{3}}{dI_{\nu}^{+} \over dz} 
&=&\ -(\kappa_{\nu} +\sigma)I_{\nu}^{+}+\sigma J_{\nu}+\epsilon_{\nu} 
\\ \nonumber {-1 \over \sqrt{3}}{dI_{\nu}^{-} 
\over dz} &=& -(\kappa_{\nu} +\sigma)I_{\nu}^{-}+\sigma J_{\nu}+\epsilon_{\nu} 
\label{eq-transA1}
\end{eqnarray}
where $I_{\nu}^{+}$ and $I_{\nu}^{-}$ are respectively the outward and inward
 intensities. 

During a Compton diffusion, a photon with an energy $E$ can lose an 
average energy $<\Delta E_{dc}>= E^2/mc^2$ for a direct 
Compton scattering, and gain an average energy $<\Delta  E_{ic}>$ = 
$E\times 4kT/mc^2$  for an inverse
Compton scattering. It is can be
sent outside the Doppler core, provided that 
$\vert \Delta E_{dc} - \Delta E_{ic}\vert \ge cst\ \Delta \nu_{D}$, where 
$cst$ is 
of the order of unity, and $\Delta \nu_{D}$ is the Doppler width. It 
is then no longer absorbed inside the line. 

In the transfer equation, we therefore treat differently the 
diffusion term $\sigma J_{\nu}$ for those photons which satisfy the previous 
condition (actually the value of $cst$ is not important): 
they are transferred like continuum photons. Their mean intensity is computed as:
\begin{equation}
J_{Compton}={\sigma \int J_{\nu} d\nu \over \Delta \nu_{Compton}},
\label{Jcomp}
\end{equation}
where $\Delta\nu_{Compton}$ is set equal to: 
\begin{equation}
\sqrt{\pi}\nu\sqrt{{2kT(0)\over mc^2} + {2\over 5}\left({h\nu\over 
mc^2}\right)^{2}}
\label{dnucomp}
\end{equation}
where $T(0)$ is the surface temperature (actually $T$ is almost constant in 
the hot Comptonizing medium). 
These photons are taken into account like continuum photons 
in the heating and ionization equations. Conversely, continuum photons are 
injected in the line core by 
Compton scattering.

Comptonization in the lines is therefore now taken into account in the statistical 
equilibrium equations, in 
the transfer, and in the emerging line fluxes.  
The gains or losses corresponding to the Compton shift of line 
photons were already taken into account in the energy balance in the previous version of 
Titan.

\bigskip

\noindent {\bf APPENDIX B: Some details concerning Escape 14bis approximation}

The local line cooling function is taken as:
\begin{equation}
\Lambda_{line}={n_{2}A_{21}h\nu\over n_{e}n_{H}} \times P_{escl}.
\label{eq-lambda-line}
\end{equation}

The reflected flux is computed as:
\begin{equation}
F_{ref}={1\over 2} \int{n_{2}A_{21}h\nu \beta'_{ref}\ exp[-\tau_{e}] dz},
\label{eq-Fref}
\end{equation}
 and the outward emitted flux is 
computed as:
\begin{equation}
F_{out}={1\over 2} \int{n_{2}A_{21}h\nu \beta'_{out}\ 
exp[-T_{e}+\tau_{e})] dz},
\label{eq-Fout}
\end{equation}
where $T_{e}$ is the total optical thickness of the slab.
 $\beta'_{ref}= P_{line}(\tau)+{1\over 2} (1-P_{escl}(\tau)) {\sigma\over 
\kappa_l\sqrt{\pi}+\kappa_c+\sigma}$, and $\beta'_{out}= P_{line}(T-\tau)+
{1\over 2} (1-P_{escl}(\tau)) {\sigma\over 
\kappa_l\sqrt{\pi}+\kappa_c+\sigma}$.
$\tau_{e}$ is the effective continuum optical thicknesses 
in the two-stream approximation 
($\tau_{e}=\sqrt{3\tau_{abs}(\tau_{abs}+\tau_{dif})}$). 

The ionization rate $\kappa_c 4\pi J_l/ h\nu$ 
due to the lines, at the depth $z$, is equal to: 

\begin{eqnarray}
&&\kappa_c A_{21}\times {\bf [}  \int_0^z{n_{2}(Z) 
\beta'_{out}(Z)exp(-\tau_{e}(z)+\tau_{e}(Z))dZ}
\\
\nonumber
&&+ \int_H^z{n_{2}(Z) 
\beta'_{ref}(Z)exp(-\tau_{e}(H)+\tau_{e}(Z))dZ}{\bf ]} ,
\label{eq-ion}
\end{eqnarray}
where $H$ is the geometrical thickness of the slab. 
This expression is used for the gains by photoionizations due to the 
lines. 



\end{document}